\pdfoutput=1
\documentclass{SCIS2026}

\usepackage{etoolbox}
\makeatletter

\patchcmd{\@maketitle}{\footnote{*{\thinspace}Corresponding author (email:~\@authoremail)}}{}{}{}
\patchcmd{\@maketitle}{\footnote{*{\thinspace}Corresponding author (email:~\@authoremail)}}{}{}{}

\patchcmd{\@maketitle}{\put(185,23){\journalnamesize\bf \textcolor[rgb]{0.01,0.12,0.67}{SCIENCE CHINA}}}{}{}{}
\patchcmd{\@maketitle}{\put(187,8){\journalnamesize \textcolor[rgb]{0.01,0.12,0.67}{Information Sciences}}}{}{}{}
\def\citationline{\@titlecitation. Preprint, 2026}
\makeatother

\makeatletter
\def\@addressline#1#2{\@addresscr\hbox{\textsuperscript{\rm#1}\kern0.15em}#2}
\makeatother

\usepackage{tikz}
\usetikzlibrary{fit,backgrounds,arrows.meta,positioning,calc}
\graphicspath{{figures/}}

\usepackage{pifont}
\usepackage{multirow}
\newcommand{\cmark}{\textcolor{green!60!black}{\ding{51}}}
\newcommand{\xmark}{\textcolor{red}{\ding{55}}}
\newcommand{\pmark}{\textcolor{orange!90!black}{\ding{115}}}  

\newsavebox{\lmleftbox}

\definecolor{modtext}{RGB}{0,112,255}   
\definecolor{modimg}{RGB}{34,139,34}    
\definecolor{modaud}{RGB}{204,102,0}    
\definecolor{modvid}{RGB}{142,68,173}   
\newcommand{\mText}{\textcolor{modtext}{Text}}
\newcommand{\mImage}{\textcolor{modimg}{Image}}
\newcommand{\mAudio}{\textcolor{modaud}{Audio}}
\newcommand{\mVideo}{\textcolor{modvid}{Video}}

\begin{document}

\ArticleType{REVIEW}
\Year{2025}
\Month{January}
\Vol{68}
\No{1}
\DOI{}
\ArtNo{}
\ReceiveDate{}
\ReviseDate{}
\AcceptDate{}
\OnlineDate{}
\AuthorMark{Yu MA, et al.}
\AuthorCitation{Ma Y, Gao Z, Qiao L, et al.}

\title{From Semantic to Token Communication: The Next Paradigm for Large-Model-Driven 6G Intelligent Connectivity}{From Semantic to Token Communication}

\author[1,2]{Yu MA}{}
\author[1]{Zhen GAO}{}
\author[3]{Li QIAO}{}
\author[2]{Xiaoyuan ZHANG}{}
\author[4]{\\Mahdi Boloursaz MASHHADI}{}
\author[5]{Yin XU}{}
\author[6]{Wenjun XU}{}
\author[6]{Xiaodong XU}{}
\author[3]{\\Kaibin HUANG}{}
\author[7]{Jiangzhou WANG}{}
\author[4]{Rahim TAFAZOLLI}{}
\author[8,9]{Sheng CHEN}{}
\author[10]{\\Tony Q. S. QUEK}{}
\author[6]{Ping ZHANG}{}

\address[1]{Beijing Institute of Technology, Beijing 100081, China}
\address[2]{Zhongguancun Academy, Beijing 100094, China}
\address[3]{Department of Electrical and Computer Engineering, The University of Hong Kong, Pokfulam Road, Hong Kong}
\address[4]{5G/6G Innovation Centre (5G/6GIC), Institute for Communication Systems,\\ University of Surrey, Guildford GU2 7XH, UK}
\address[5]{Cooperative Medianet Innovation Center, Shanghai Jiao Tong University, Shanghai 200240, China}
\address[6]{State Key Laboratory of Networking and Switching Technology,\\ Beijing University of Posts and Telecommunications, Beijing 100876, China}
\address[7]{School of Information Science and Engineering, Southeast University, Nanjing 210096, China}
\address[8]{School of Electronics and Computer Science, University of Southampton, Southampton SO17 1BJ, UK}
\address[9]{Faculty of Information Science and Technology, Ocean University of China, Qingdao 266100, China}
\address[10]{Information Systems Technology and Design Pillar, Singapore University of Technology and Design, Singapore 487372, Singapore}

\abstract{%
The ambitious requirements of sixth-generation (6G) networks are driving communication systems from reliable bit delivery toward meaning-aware and task-oriented connectivity. Large models (LMs), with strong multimodal understanding and generation capabilities, have accelerated this shift and made semantic communication (SemCom) increasingly practical. Yet current LM-driven SemCom remains fragmented: semantic representations are typically tied to specific modalities, models, or tasks. While the bit provides a universal unit for digital transport, there is still no analogous unit for representing and processing semantics, which limits interoperability, theoretical unification, and scalable system design.
We argue that tokens provide a natural candidate for this missing abstraction. Two trends support this: unified multimodal LMs now encode text, images, audio, video, and robot actions in one token space, while distributed LM inference already generates substantial token-level traffic through expert routing, cache transfer, and speculative decoding. Token communication (TokenCom) emerges by unifying these trends, using the LM's native processing unit as a communication abstraction above the bit level and enabling importance assignment, error handling, and resource allocation directly at token granularity.
This survey traces the evolution from LM-driven SemCom to TokenCom. We review three major directions of LM-driven SemCom: source-centric semantic coding, channel semantics for physical-layer tasks, and collaborative edge-device intelligence. We then examine the token abstraction, the transmission techniques it requires, and two emerging paradigms, namely TokenCom for LM services and for embodied and agentic intelligence. Finally, we identify open challenges toward unified, scalable, and AI-native 6G communication systems.}


\maketitle

\section{Introduction}\label{sec:intro} 

Sixth-generation (6G) networks are expected to support immersive and intelligent services such as extended reality, metaverse applications, autonomous systems, and ubiquitous machine interaction. To this end, the International Telecommunication Union (ITU) has identified the integration of artificial intelligence (AI) and communication as a representative 6G usage scenario~\cite{recommendation2023framework}. This vision reflects a broader transition from data-centric transmission to intelligent connectivity~\cite{cuiOverviewAICommunication2025}, in which communication systems are designed not only to deliver bits reliably, but also to support semantic understanding, reasoning, and decision-making.

Such a transition is driven by stringent performance requirements. Emerging 6G services demand extreme traffic density, ultra-high data rates, and sub-millisecond latency, that is, traffic density on the order of $1$--$10$~$\mathrm{Gbps/m^3}$, peak rates up to $1$~$\mathrm{Tbps}$, and latency as low as $0.1$~$\mathrm{ms}$~\cite{CALVANESESTRINATI2021107930}. Conventional communication systems are fundamentally built upon Shannon's syntactic communication theory~\cite{6773024}, where the primary objective is accurate symbol reconstruction. Although modern coding and modulation techniques have approached Shannon capacity, bit-level transmission alone is increasingly insufficient for supporting meaning-aware and task-oriented intelligent services.

Semantic communication (SemCom) has therefore emerged as a promising alternative. Rather than treating all bits equally, SemCom prioritizes the delivery of meaning or task-relevant information, thereby reducing unnecessary transmission overhead and improving communication efficiency. Rooted in the semantic level of communication envisioned by Shannon and Weaver~\cite{weaverRecentContributionsMathematical}, SemCom shifts the design objective from symbol fidelity to semantic fidelity. This paradigm is particularly attractive for 6G systems, where resource efficiency, robustness, and intelligent interaction are all critical.

A key challenge, however, is that SemCom requires models capable of extracting, representing, and reconstructing semantics across diverse data modalities and tasks. Recent advances in deep learning, especially large models (LMs), have made this feasible. Transformer-based autoregressive (AR) models have demonstrated strong capabilities in semantic understanding and generation for language, and these capabilities are now extending to multimodal data such as images, audio, and video~\cite{gpt3.5,wangMultimodalLearningNexttoken2026,luUnifiedIO2Scaling2024}. At the same time, diffusion and related generative architectures enable high-quality content reconstruction, making LMs increasingly suitable as the computational core of communication systems.

From a communication perspective, LMs provide an important new capability: through large-scale pretraining, transceivers can share implicit prior knowledge and use it to reconstruct information from compact semantic cues rather than from fully recovered symbol streams. This property reduces the reliance on strict bit-wise fidelity and enables semantic-aware communication strategies with substantially improved efficiency. Nevertheless, important challenges remain. Inference with LMs introduces significant latency and computational cost, especially on edge devices. More fundamentally, semantic theory still lacks mature mathematical foundations and quantitative metrics~\cite{chafiiTwelveScientificChallenges2023}, and it lacks, we argue, a unified representation unit comparable to the bit, leaving semantic representations tied to specific modalities and models. This fragmentation limits interoperability, theoretical unification, and scalable system design.

Recent progress in multimodal LMs suggests a compelling direction to address this issue. In particular, token-based modeling has shown that text, images, audio, and video can be represented and processed within a unified discrete interface~\cite{wangMultimodalLearningNexttoken2026,luUnifiedIO2Scaling2024}. Practical systems such as GPT-4o~\cite{openaiGPT4oSystemCard2024} and Gemini~\cite{reidGemini15Unlocking2024} further demonstrate that tokens can support coherent multimodal understanding and generation in a single framework. These developments suggest that tokens may serve as a universal information unit for future intelligent communication systems.

\begin{figure}
    \centering
    \includegraphics[width=0.9\textwidth]{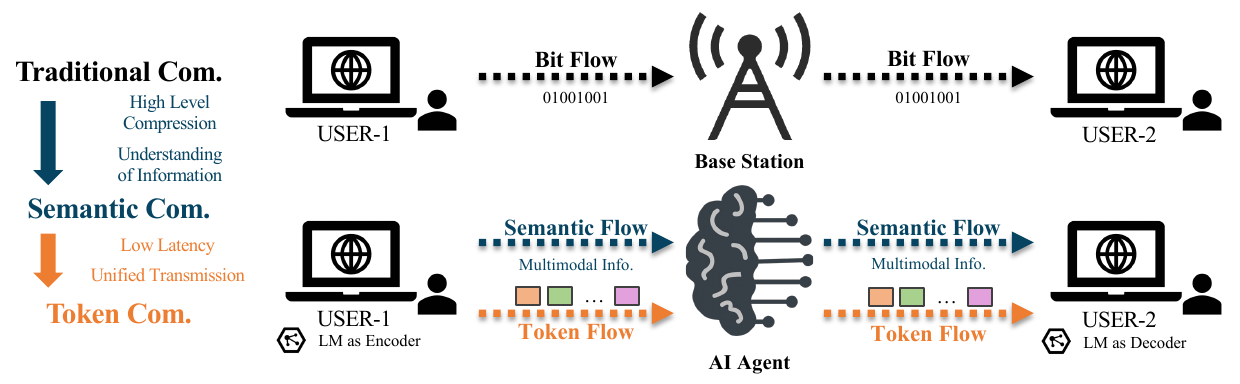}
    \caption{Evolution from bit-level communication to semantic-level communication and further to token-level communication enabled by LMs. The three flows label the unit at which each paradigm represents and optimizes information. All three still require physical transmission, in either digital or analog form.}
    \label{fig:intro} 
\end{figure}

Motivated by this observation, token communication (TokenCom) has recently been proposed as an evolution of SemCom~\cite{qiaoTokenCommunicationsLarge2025}. Instead of relying on abstract and heterogeneous semantic representations, TokenCom adopts tokens, which serve as the native interface of LMs, as the fundamental communication unit. Because tokens provide a structured and modality-agnostic representation aligned with model inference, token-level communication offers a natural way to connect data, inference, and transmission in a unified framework. As illustrated in Figure~\ref{fig:intro}, this perspective represents a natural evolution from bit-level communication to semantic-level communication, and further to token-level communication for intelligent connectivity.

To provide a structured overview, this survey covers LM-driven SemCom along three established directions and TokenCom through the token abstraction, the transmission techniques it requires, and two emerging paradigms, with the directions and paradigms characterized as follows.

\textbf{Source-Centric Semantic Coding}: This direction exploits the generative capabilities of LMs to perform meaning-aware compression and reconstruction. By extracting high-dimensional semantic features, these approaches shift the reconstruction target from bit-level fidelity to semantic-level synthesis, moving from separate source-channel coding (SSCC) toward joint source-channel coding (JSCC) to improve robustness~\cite{qiaoLatencyAwareGenerativeSemantic2024,11112664,jiangLargeAIModelBased2023}.

\textbf{Channel Semantics for Physical-Layer Tasks}: Unlike conventional physical-layer techniques, this direction treats channel state information (CSI) as a semantic source. Task-specific deep networks first showed that CSI can be compressed into compact channel semantics for pilot design, feedback, and beamforming~\cite{gaoHybridKnowledgeDataDriven2023,zhangAIEmpoweredChannel2024}. LMs extend this line by handling several such tasks within a single model.

\textbf{Collaborative Edge-Device Intelligence}: Recognizing the high computational overhead of LMs, this direction focuses on distributed architectures. It explores the synergy between resource-constrained devices and edge servers~\cite{zhengCommunicationEfficientCollaborativeLLM2025}, investigating model selection and offloading and collaborative inference to facilitate the practical deployment of LMs in wireless networks~\cite{renGenerativeSemanticCommunication2025}.

\textbf{Token Communication for LM Services}: This emerging paradigm broadens the scope of communication design from serving semantic fidelity to directly enabling LM-powered AI services. In distributed LM deployments, token transmission is inherently embedded in the inference pipeline: the transfer of key-value (KV) cache states synchronizes LM state across nodes. Mixture-of-experts (MoE) routing dispatches tokens to expert modules residing on different network nodes, and distributed speculative decoding shuttles draft and verification tokens between edge devices and servers. Efficient token scheduling and delivery thus become essential enablers of end-to-end LM service quality, motivating a communication paradigm co-designed with AI inference rather than merely supporting data delivery.

\textbf{Token Communication for Embodied and Agentic Intelligence}: In this paradigm, at least one endpoint of the link is an agent that acts on its physical environment, so that perception, communication, and action form a closed loop. Transmission quality is then measured by task execution and control stability rather than by reconstruction fidelity. Actions and observations can be carried over a single token interface. A transmission is triggered by what the receiving agent still needs in order to act rather than by what the channel can currently carry.

Although a growing body of survey literature has reviewed SemCom from the perspectives of theory, system architecture, applications, and generative AI~\cite{iyerSurveySemanticCommunications2023,getuSemanticCommunicationSurvey2024,guoSurveySemanticCommunication2024a,liangGenerativeAIDrivenSemantic2025,xiaGenerativeAISemantic2025,aloudatMetaverseUnboundSurvey2025,chenSurveySemanticExtraction2025,nguyenContemporarySurveySemantic2025,renGenerativeSemanticCommunication2025a,yingSpecialistLargeModels2026}, two gaps remain. First, existing surveys generally treat LMs as enabling tools rather than as the core intelligence substrate of next-generation communication systems. Second, the role of tokens as a unified representation unit has not been systematically examined. As a result, the transition from SemCom to token-based communication remains insufficiently articulated. Table~\ref{tab:survey_comparison} contrasts this survey with the existing ones along the five research directions and paradigms identified above: while source-centric semantic coding is extensively surveyed, channel semantics for physical-layer tasks receives only scattered attention, and the token-centric perspective as well as the two emerging paradigms are substantively covered by none of the existing surveys.

\begin{table*}[t]
\centering
\caption{Comparison of this survey with existing SemCom surveys, aligned with the three research directions and two emerging paradigms identified in this paper. Foundations --- LM: large-model fundamentals; Token: token as the communication unit. Research directions and paradigms --- Sem. coding: source-centric semantic coding; Ch. semantics: channel semantics for physical-layer tasks; Edge collab.: collaborative edge-device intelligence; LM serv.: token communication for LM services; Embodied: embodied and agentic intelligence (\cmark: substantive coverage; \pmark: partial; \xmark: not covered)}
\label{tab:survey_comparison} 
\small
\setlength{\tabcolsep}{6pt}
\begin{tabular}{|l|c|cc|ccccc|}
\hline
 & & \multicolumn{2}{c|}{\textbf{Foundations}} & \multicolumn{5}{c|}{\textbf{Research directions \& paradigms}} \\
\textbf{Reference} & \textbf{Year} & LM & Token & Sem. coding & Ch. semantics & Edge collab. & LM serv. & Embodied \\
\hline
Iyer et al.~\cite{iyerSurveySemanticCommunications2023} & 2023 & \xmark & \xmark & \cmark & \xmark & \pmark & \xmark & \xmark \\
Getu et al.~\cite{getuSemanticCommunicationSurvey2024} & 2024 & \xmark & \xmark & \cmark & \pmark & \pmark & \xmark & \xmark \\
Guo et al.~\cite{guoSurveySemanticCommunication2024a} & 2025 & \xmark & \xmark & \pmark & \xmark & \pmark & \xmark & \pmark \\
Liang et al.~\cite{liangGenerativeAIDrivenSemantic2025} & 2025 & \cmark & \xmark & \cmark & \xmark & \pmark & \pmark & \xmark \\
Xia et al.~\cite{xiaGenerativeAISemantic2025} & 2025 & \pmark & \xmark & \cmark & \xmark & \cmark & \pmark & \xmark \\
Aloudat et al.~\cite{aloudatMetaverseUnboundSurvey2025} & 2025 & \pmark & \xmark & \pmark & \xmark & \cmark & \xmark & \xmark \\
Chen et al.~\cite{chenSurveySemanticExtraction2025} & 2025 & \pmark & \xmark & \pmark & \xmark & \xmark & \xmark & \xmark \\
Nguyen et al.~\cite{nguyenContemporarySurveySemantic2025} & 2026 & \pmark & \xmark & \cmark & \xmark & \pmark & \xmark & \pmark \\
Ren et al.~\cite{renGenerativeSemanticCommunication2025a} & 2026 & \cmark & \pmark & \cmark & \xmark & \pmark & \xmark & \pmark \\
Ying et al.~\cite{yingSpecialistLargeModels2026} & 2026 & \pmark & \pmark & \pmark & \cmark & \xmark & \xmark & \xmark \\
\hline
\textbf{This survey} & 2026 & \cmark & \cmark & \cmark & \cmark & \cmark & \cmark & \cmark \\
\hline
\end{tabular}
\end{table*}

This paper addresses these gaps by providing a unified review of LM-driven SemCom from a token-centric perspective. We argue that LMs should be viewed as the intelligence engine of future communication systems, while tokens should be regarded as a candidate fundamental unit for multimodal semantic information exchange. Under this perspective, SemCom and TokenCom can be understood within a common framework, which helps clarify current progress, expose open challenges, and identify future research opportunities.

The main contributions of this paper are summarized as follows:
\begin{enumerate}
    \item We review the fundamentals of LMs in the context of communication systems and clarify why tokens constitute a natural semantic interface for multimodal information processing.
    \item We provide a systematic taxonomy of LM-driven SemCom, characterizing the transition from source-centric bit-fidelity objectives to LM-driven semantic exchange. We re-evaluate existing research through a token-centric lens, offering a unified perspective on how LMs transform source and channel processing.
    \item We identify the limitations of existing SemCom frameworks, especially the lack of a unified representation unit, and discuss TokenCom as a promising direction toward a scalable and model-aware communication paradigm.
\end{enumerate}

\begin{figure*}[t]
    \centering
    \resizebox{0.82\textwidth}{!}{\input{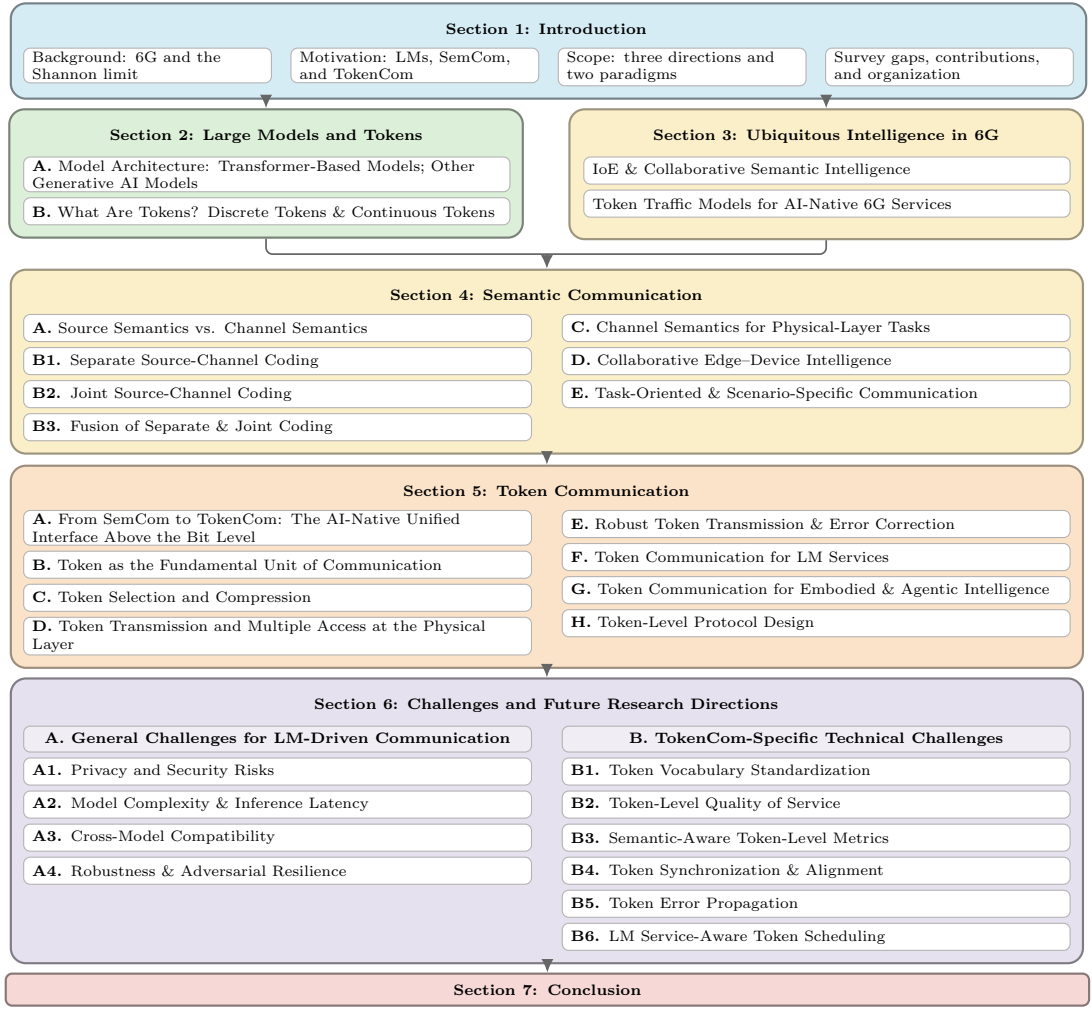}}
    \caption{Organization of this survey paper}
    \label{fig:structure} 
\end{figure*}

The remainder of this paper is structured to provide a logical flow from theoretical foundations to future paradigms. Sections~\ref{sec:LargeModel} and \ref{sec:Intelligence} establish the necessary background on LMs and tokens, and the evolving role of intelligence in networks. Sections~\ref{sec:Semcom} and \ref{sec:TokenCom} provide a comparative analysis of existing SemCom and the emerging TokenCom framework, respectively. Finally, Section~\ref{sec:Challenges} discusses open problems, followed by concluding remarks in Section~\ref{sec:Conclusion}. The overall organization of this survey is illustrated in Figure~\ref{fig:structure}.

\section{Large Models and Tokens}\label{sec:LargeModel} 

In this section, we review key concepts related to LMs, including their architectures, tokenization mechanisms, and emergent abilities. The technologies discussed here are representative, as many SemCom systems build upon the models and architectures presented below. This overview also provides the necessary background for understanding why tokens are suitable as the fundamental unit of communication.

\subsection{Model Architecture} 

LMs refer to deep learning models with millions or even billions of parameters. By pretraining on large-scale datasets, they capture complex patterns and relationships, with their success governed by neural scaling laws~\cite{kaplan2020scaling}. Several key factors underpin their effectiveness.
\begin{itemize}
    \item \textbf{Massive Data}: LMs are trained on vast corpora that provide rich semantic information, enabling them to acquire broad world knowledge.
    \item \textbf{Advanced Architectures}: Architectures such as the transformer efficiently model long-range dependencies and complex patterns through parallelized computation.
    \item \textbf{Pre-training and Fine-Tuning}: A two-stage paradigm in which models first learn general-purpose representations and are subsequently fine-tuned for specific downstream tasks.
\end{itemize}

In SemCom systems, LMs fulfill two complementary functional roles. \emph{Representation models} serve as semantic encoders, extracting compact and informative representations from multimodal inputs. \emph{Generative models} serve as semantic decoders, reconstructing or synthesizing content from these representations. Notably, transformer-based architectures contribute to both roles: encoder-style models such as vision transformer (ViT)~\cite{Dosovitskiy2020AnII}, contrastive language-image pre-training (CLIP)~\cite{radford2021learningtransferablevisualmodels}, and bidirectional encoder representation from transformers (BERT)~\cite{devlinBERTPretrainingDeep2019} provide powerful semantic representations, while decoder-style transformers support generation through either autoregressive strategies, as in the generative pre-trained transformer (GPT), or masked prediction, as in the masked generative image transformer (MaskGIT). Beyond transformers, other generative paradigms, including generative adversarial networks (GANs), variational autoencoders (VAEs), and diffusion models, are also widely employed for high-fidelity content recovery. We describe each category in detail below.

\subsubsection{Transformer-Based Models}\label{sususec:llm} 

Introduced by Vaswani et al.~\cite{vaswaniAttentionAllYou2017} in 2017, the transformer has become a foundational architecture in natural language processing (NLP) and has since been extended to other domains such as computer vision (CV). Its ability to model global dependencies and to parallelize computation has enabled scalable training. Beyond a certain scale, pre-trained LMs have been reported to show emergent abilities, that is, abilities that are not present in smaller models~\cite{wei2022emergent}.

\begin{figure*}[!t]
    \centering
    \sbox{\lmleftbox}{\begin{minipage}[b]{0.52\textwidth}
        \centering
        \subfloat[]{%
            \includegraphics[trim=0 13 0 22,clip,width=\textwidth]{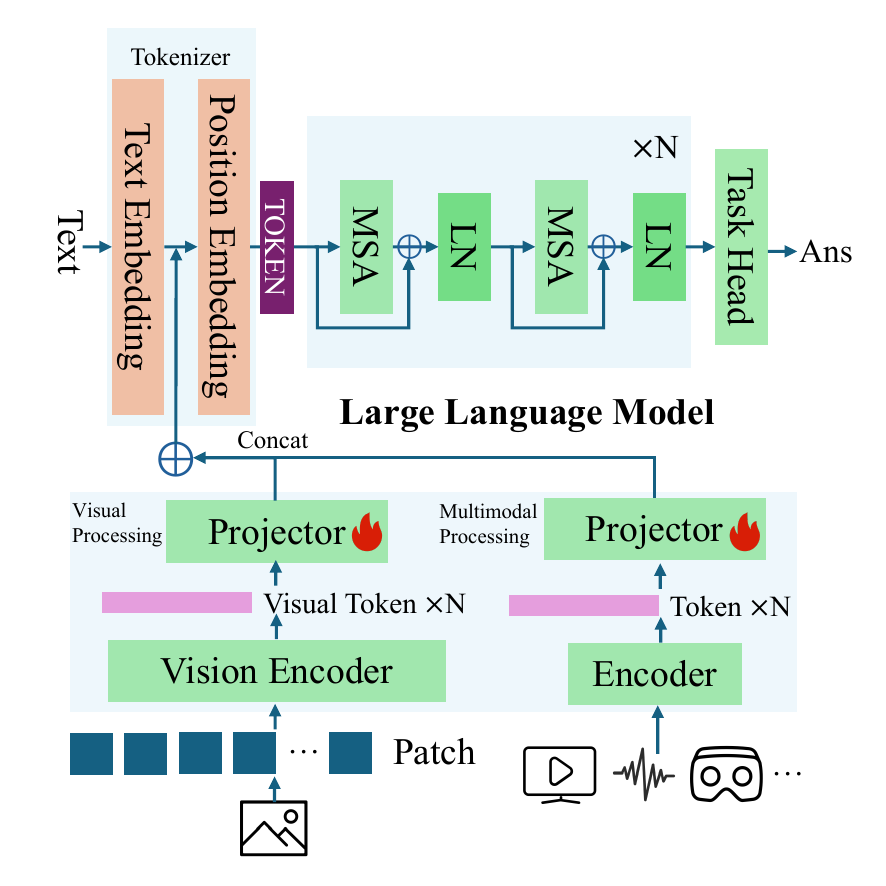}%
            \label{fig:LM_a}}
    \end{minipage}}%
    \usebox{\lmleftbox}%
    \begin{minipage}[b][\ht\lmleftbox][s]{0.42\textwidth}
        \centering
        \subfloat[]{%
            \includegraphics[trim=0 0 0 4,clip,width=\textwidth]{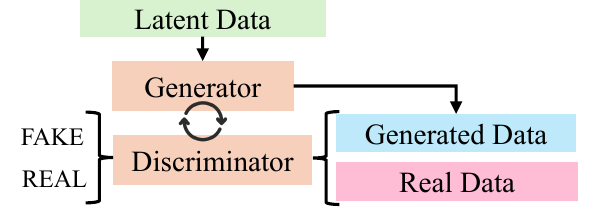}%
            \label{fig:LM_b}}\par\vfill
        \subfloat[]{%
            \includegraphics[trim=0 13 0 15,clip,width=\textwidth]{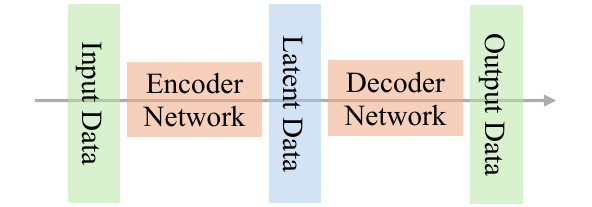}%
            \label{fig:LM_c}}\par\vfill
        \subfloat[]{%
            \includegraphics[trim=0 8 0 6,clip,width=0.75\textwidth]{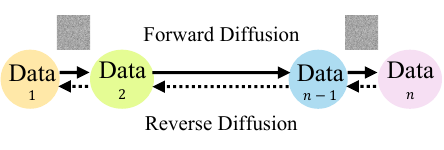}%
            \label{fig:LM_d}}
    \end{minipage}
    \caption{(a) Multimodal large language model (MSA denotes multi-head self-attention and LN denotes layer normalization), and other generative AI models, including (b) generative adversarial network, (c) variational autoencoder, and (d) diffusion probabilistic model.}
    \label{fig:LM} 
\end{figure*}

At the core of the transformer, the self-attention mechanism dynamically attends to different parts of the input sequence, producing weighted, context-aware representations for each position. In this subsection, we focus on the information flow of the self-attention mechanism, abstracting away implementation details.

A standard transformer consists of stacked encoder and decoder layers, in which the self-attention mechanism serves as the primary vehicle for information interaction. The encoder processes the input sequence to produce contextual representations, while the decoder generates the target sequence conditioned on both the encoder outputs and previously generated tokens.

Given an input sequence, each word is first tokenized into subword units (referred to as tokens) and mapped to a numerical index, yielding a token sequence $\mathbf{x} = (x_1, x_2, \dots, x_n)$ of length $n$. These tokens are then embedded into a high-dimensional space and combined with positional encodings to form the input representations $\mathbf{e}_i \in \mathbb{R}^{d}$, where $d$ is the model dimension. This embedding process captures both semantic meaning and positional information, providing the foundation for subsequent self-attention operations.

These vectors are then transformed into queries $\mathbf{q}_i$, keys $\mathbf{k}_i$, and values $\mathbf{v}_i$ through learned linear projections
\begin{equation} 
    \mathbf{q}_i = \mathbf{W}^Q \mathbf{e}_i \in \mathbb{R}^{d_k}, \mathbf{k}_i = \mathbf{W}^K \mathbf{e}_i \in \mathbb{R}^{d_k}, \mathbf{v}_i = \mathbf{W}^V \mathbf{e}_i \in \mathbb{R}^{d_v},
\end{equation}
where $d_k$ and $d_v$ denote the query/key and value dimensions, and $\mathbf{W}^Q, \mathbf{W}^K \in \mathbb{R}^{d_k \times d}$, $\mathbf{W}^V \in \mathbb{R}^{d_v \times d}$ are learned projection matrices. Stacking these vectors across all $n$ tokens yields the matrices $\mathbf{Q} = [\mathbf{q}_1^T; \mathbf{q}_2^T; \dots; \mathbf{q}_n^T]$, $\mathbf{K} = [\mathbf{k}_1^T; \mathbf{k}_2^T; \dots; \mathbf{k}_n^T]$, and $\mathbf{V} = [\mathbf{v}_1^T; \mathbf{v}_2^T; \dots; \mathbf{v}_n^T]$, where we have $\mathbf{Q}, \mathbf{K} \in \mathbb{R}^{n \times d_k}$ and $\mathbf{V} \in \mathbb{R}^{n \times d_v}$.

The dot product $\mathbf{q}_i^T \mathbf{k}_j$ quantifies the relevance between the $i$-th query and the $j$-th key, while the dot-product (score) matrix $\mathbf{Q}\mathbf{K}^T$ encodes pairwise interactions among all tokens. After normalization via the softmax function, the resulting attention weights are used to compute a weighted sum of value vectors, enabling each token to selectively aggregate information from other positions. From this perspective, self-attention acts as a dynamic information filtering mechanism that amplifies task-relevant features.
The self-attention computation is formulated as
\begin{equation}
    \text{Attention}(\mathbf{Q}, \mathbf{K}, \mathbf{V}) = \text{softmax}\left( \frac{\mathbf{Q} \mathbf{K}^T}{\sqrt{d_k}} \right) \mathbf{V}.
\end{equation} 
Although a single self-attention operation enables each token to aggregate information from all positions, it operates within a single representation subspace. To enhance expressiveness, the transformer employs multi-head self-attention (MSA), which allows the model to capture different types of relationships in parallel.

Specifically, writing $\mathbf{E} = [\mathbf{e}_1^T; \dots; \mathbf{e}_n^T] \in \mathbb{R}^{n \times d}$ for the stacked input representations, the $m$-th of the $h$ heads applies its own projections to $\mathbf{E}$ to obtain $\mathbf{Q}_m$, $\mathbf{K}_m$, and $\mathbf{V}_m$, so that its output reads
\begin{equation} 
    \text{head}_{m} = \text{Attention}(\mathbf{Q}_{m},\,
    \mathbf{K}_{m},\,
    \mathbf{V}_{m}).
\end{equation}
The outputs of all heads are concatenated along the feature dimension and linearly projected to obtain the multi-head attention output
\begin{equation} 
    \text{MSA}(\mathbf{E}) = \text{Concat}(\text{head}_1, \ldots, \text{head}_h) \mathbf{W}^O,
\end{equation}
where $\mathbf{W}^O \in \mathbb{R}^{h d_v \times d}$ denotes the learnable output projection matrix, with $d_k = d_v = d/h$ the setting commonly adopted. By jointly attending to multiple representation subspaces, multi-head attention improves the model's representational capacity and enables more effective modeling of complex dependencies.
In contrast to the \emph{encoder-decoder} architecture, modern LMs such as GPT adopt a \emph{decoder-only} transformer architecture, as illustrated in Figure~\ref{fig:LM_a}. Given an input sequence, the model generates tokens autoregressively, predicting one token at a time conditioned on all previously generated tokens. Open-source models such as LLaMA~\cite{touvronLLaMAOpenEfficient2023} follow a similar architectural paradigm, differing primarily in implementation details such as positional encoding schemes and activation functions.

Given a tokenized input sequence $\mathbf{x} = (x_1, x_2, \dots, x_n)$, the model first maps each token to a continuous embedding and incorporates positional information
\begin{equation} 
    \mathbf{h}^{[0]} = \mathrm{Embed}(\mathbf{x}) + \mathbf{p},
\end{equation}
where $\mathrm{Embed}(\cdot)$ denotes the token embedding function, $\mathbf{p} \in \mathbb{R}^{n \times d}$ represents the positional encoding, and $\mathbf{h}^{[0]} \in \mathbb{R}^{n \times d}$ is the initial hidden representation.

The embeddings are then processed by a stack of $L$ decoder layers. At the $\ell$-th layer, the input hidden state $\mathbf{h}^{[\ell-1]}$ is first normalized and passed through a masked MSA module
\begin{equation} 
    \mathbf{h}' = \mathbf{h}^{[\ell-1]} +
    \mathrm{MSA}\big(\mathrm{LN}(\mathbf{h}^{[\ell-1]})\big),
\end{equation}
where $\mathrm{LN}(\cdot)$ denotes layer normalization, and the masking operation ensures that each token attends only to its past context, preserving the autoregressive property.

Subsequently, a position-wise feed-forward network (FFN) further transforms the representations
\begin{equation} 
    \mathbf{h}^{[\ell]} = \mathbf{h}' +
    \mathrm{FFN}\big(\mathrm{LN}(\mathbf{h}')\big),
    \label{eq:decoder_ffn}
\end{equation}
where $\mathbf{h}^{[\ell]} \in \mathbb{R}^{n \times d}$ is the output of the $\ell$-th decoder layer. Residual connections are employed in both sublayers to facilitate stable training.

After the final decoder layer, the hidden state corresponding to each position is projected to a vocabulary-sized logit vector to predict the next token
\begin{equation} 
    \hat{\mathbf{Y}} = \mathrm{softmax}\big(\mathbf{h}^{[L]} \mathbf{W}_o^{T}\big) \in \mathbb{R}^{n \times |\mathcal{V}|},
\end{equation}
where $\mathbf{h}^{[L]}$ denotes the output of the final ($L$-th) decoder layer, obtained by iterating Eq.~(\ref{eq:decoder_ffn}) up to $\ell = L$, $\mathbf{W}_o \in \mathbb{R}^{|\mathcal{V}| \times d}$ is the output projection matrix, $|\mathcal{V}|$ denotes the vocabulary size, and $\hat{\mathbf{Y}}$ collects the predicted token distributions over all $n$ positions, with the softmax applied row-wise.

Building upon the success of decoder-only transformers in NLP, this architecture has been extended to multimodal systems such as LLaVA~\cite{liu2023llava}, which integrate visual and linguistic information within a unified framework. In such models, different modalities are first processed by modality-specific encoders and then projected into a shared token representation space.

For the visual modality, a pre-trained vision encoder (e.g., CLIP~\cite{radford2021learningtransferablevisualmodels} with a ViT backbone~\cite{Dosovitskiy2020AnII}) extracts feature representations from an input image $\mathbf{I}$
\begin{equation} 
    \mathbf{f}_I = \mathrm{VisionEncoder}(\mathbf{I}),
\end{equation}
where $\mathbf{f}_I$ denotes the high-level visual features.

These visual features are then projected into the language model's token embedding space
\begin{equation} 
    \mathbf{z}_I = \mathrm{Projector}(\mathbf{f}_I),
    \label{eq:projector}
\end{equation}
where $\mathbf{z}_I$ represents the resulting visual tokens aligned with the text token space.

Finally, the visual tokens $\mathbf{z}_I$ are concatenated with text tokens $\mathbf{x}$ and jointly processed by the large language model (LLM)
\begin{equation} 
    \hat{\mathbf{y}} = \mathrm{LLM}\big([\mathbf{z}_I, \mathbf{x}]\big).
\end{equation}

This unified token-based formulation enables flexible integration of heterogeneous modalities, including images, video, and audio, making transformer-based architectures particularly well-suited for SemCom systems.

\subsubsection{Other Generative AI Models} 

Autoregressive architectures such as GPT and LLaVA represent the most widely recognized class of generative models. In contrast, alternative paradigms, namely, GANs, VAEs, and diffusion models, have demonstrated superior performance in generating images, videos, and other high-dimensional continuous data. This subsection provides a concise overview of these models, as illustrated in Figures~\ref{fig:LM_b}--\ref{fig:LM_d}.

GANs~\cite{goodfellowGenerativeAdversarialNets2014}, shown in Figure~\ref{fig:LM_b}, consist of a generator $G$ and a discriminator $D$ engaged in a minimax game to approximate the true data distribution. The generator maps a random vector $z \sim p(z)$ to the data space, while the discriminator distinguishes real samples from generated ones. The training objective is
\begin{equation} 
\begin{split}
\min_{G}\max_{D}\; V(D,G) = &\; \mathbb{E}_{x\sim p_{\text{data}}(x)}[\log D(x)] + \mathbb{E}_{z\sim p(z)}[\log(1 - D(G(z)))].
\end{split}
\end{equation}
At equilibrium, the generator's output distribution converges to the real data distribution, and the discriminator can no longer distinguish between real and generated samples.

VAEs~\cite{kingmaAutoEncodingVariational2014}, shown in Figure~\ref{fig:LM_c}, employ an encoder-decoder architecture and are trained by maximizing the evidence lower bound (ELBO)
\begin{equation} 
\mathcal{L}_{\text{ELBO}} =
\mathbb{E}_{q_{\phi}(z|x)}[\ln p_{\theta}(x|z)]
- D_{\text{KL}}\big(q_{\phi}(z|x)\parallel p(z)\big).
\end{equation}
The VAE compresses data into a probabilistic latent space, from which new samples can be drawn by sampling from the prior $p(z)$.

Diffusion models, shown in Figure~\ref{fig:LM_d}, learn the data distribution through a forward process that gradually adds Gaussian noise and a reverse process that iteratively denoises. In denoising diffusion probabilistic models (DDPMs)~\cite{hoDenoisingDiffusionProbabilistic2020}, a network $\epsilon_\theta$ is trained to predict the injected noise by minimizing
\begin{equation} 
L_{\text{diff}} = \mathbb{E}_{t,x_0,\epsilon}\left[\left\|\epsilon - \epsilon_\theta(x_t,t)\right\|^2\right],
\end{equation}
and new samples are generated by iteratively denoising a Gaussian prior. Since the step-by-step reverse process is computationally expensive, denoising diffusion implicit models (DDIMs)~\cite{songDenoisingDiffusionImplicit2021} reformulate it as a non-Markovian process that supports much fewer sampling steps. Latent diffusion models (LDMs)~\cite{rombachHighResolutionImageSynthesis2022}, popularized as Stable Diffusion, further shift the diffusion process from the pixel space into the compact latent space of a pre-trained autoencoder and introduce cross-attention conditioning on text embeddings, enabling high-resolution text-to-image generation at practical cost. More recently, diffusion transformers (DiTs)~\cite{peeblesScalableDiffusionModels2023} replace the conventional U-Net denoiser with a transformer operating on latent patch tokens, showing that the favorable scaling behavior of transformers carries over to diffusion-based generation.

In SemCom systems, diffusion models have mainly been exploited as powerful semantic decoders: compact received semantics (e.g., textual prompts, sketches, or latent tokens) condition the reverse process to reconstruct perceptually faithful content under stringent bandwidth constraints~\cite{qiaoLatencyAwareGenerativeSemantic2024,guoDiffusionDrivenSemanticCommunication2024,maLGVSCLargeModelDriven2026}, while latent-diffusion-based designs further mitigate semantic ambiguities and channel noise~\cite{peiLatentDiffusionModelEnabled2025,xuSemanticPriorAided2025}.

Table~\ref{tab:models} summarizes the representative transformer-based and generative AI models discussed above in terms of architecture, modality, and token representation.

\begin{table*}[!t]
\centering
\caption{Representative transformer-based and other generative AI models, grouped by architecture family. Modalities are color-coded as \mText{}, \mImage{}, \mAudio{}, and \mVideo{}. Token type --- D: discrete; C: continuous; D+C: hybrid; n.d.: not publicly disclosed; parenthetical notes indicate the specific representation where noteworthy}
\label{tab:models} 
\footnotesize
\setlength{\tabcolsep}{3pt}
\begin{tabular}{|p{2.1cm}|l|c|p{2.8cm}|p{2.8cm}|p{4.2cm}|}
\hline
\textbf{Architecture} & \textbf{Model} & \textbf{Year} & \textbf{Modality} & \textbf{Token type} & \textbf{Key characteristic} \\
\hline
Encoder--decoder & Transformer~\cite{vaswaniAttentionAllYou2017} & 2017 & \mText & D & Replaces recurrence with parallelizable attention \\
\hline
\multirow{2}{=}{Encoder-only} & BERT~\cite{devlinBERTPretrainingDeep2019} & 2019 & \mText & D & Bidirectional masked-LM pretraining \\
\cline{2-6}
 & ViT~\cite{Dosovitskiy2020AnII} & 2021 & \mImage & C (patch embeddings) & Images processed as patch-token sequences \\
\hline
\multirow{5}{=}{Decoder-only (AR)} & GPT-3~\cite{brownLanguageModelsAre2020} & 2020 & \mText & D & Few-shot in-context learning at 175B scale \\
\cline{2-6}
 & LLaMA~\cite{touvronLLaMAOpenEfficient2023} & 2023 & \mText & D & Open-weight LLM family optimized for inference cost \\
\cline{2-6}
 & GPT-4o~\cite{openaiGPT4oSystemCard2024} & 2024 & \mText, \mImage, \mAudio, \mVideo & n.d. & End-to-end real-time multimodal interaction \\
\cline{2-6}
 & Gemini 1.5~\cite{reidGemini15Unlocking2024} & 2024 & \mText, \mImage, \mAudio, \mVideo & n.d. & Sparse MoE; million-token-scale long context \\
\cline{2-6}
 & Emu3~\cite{wangMultimodalLearningNexttoken2026} & 2026 & \mText, \mImage, \mVideo & D (shared vision--text vocabulary) & Next-token prediction unifies perception and generation \\
\hline
Dual-encoder (contrastive) & CLIP~\cite{radford2021learningtransferablevisualmodels} & 2021 & \mText, \mImage & C (joint embedding) & Zero-shot vision--language alignment \\
\hline
Vision encoder + LLM & LLaVA~\cite{liu2023llava} & 2023 & \mText, \mImage & D+C (text\,/\,vision) & Visual instruction tuning for multimodal LLMs \\
\hline
Adversarial & GAN~\cite{goodfellowGenerativeAdversarialNets2014} & 2014 & \mImage & C & Implicit generative modeling via adversarial game \\
\hline
Variational autoencoder & VAE~\cite{kingmaAutoEncodingVariational2014} & 2014 & \mImage & C & ELBO-based variational inference \\
\hline
\multirow{3}{=}{Diffusion} & DDPM~\cite{hoDenoisingDiffusionProbabilistic2020} & 2020 & \mImage & C (pixel space) & Iterative denoising rivals GANs in fidelity \\
\cline{2-6}
 & LDM~\cite{rombachHighResolutionImageSynthesis2022} & 2022 & \mText\,$\to$\,\mImage & C (VAE latent) & High-resolution synthesis in compact latent space \\
\cline{2-6}
 & DiT~\cite{peeblesScalableDiffusionModels2023} & 2023 & \mImage & C (latent patches) & Transformer scaling behavior for diffusion \\
\hline
\end{tabular}
\end{table*}

\subsection{What Are Tokens?}\label{subsec:token} 

Tokens are the fundamental semantic units in LMs, serving as the bridge between modality-specific source signals and the unified semantic space used by these models. Depending on their representation form and processing pathway, tokens can be categorized into discrete tokens and continuous tokens, each exhibiting distinct properties and advantages in semantic representation, robustness, and integration with communication systems. A multimodal LM commonly handles both types simultaneously, as listed for LLaVA~\cite{liu2023llava} in Table~\ref{tab:models}.

\begin{figure}[!t]
    \centering
    \includegraphics[width=0.4\textwidth]{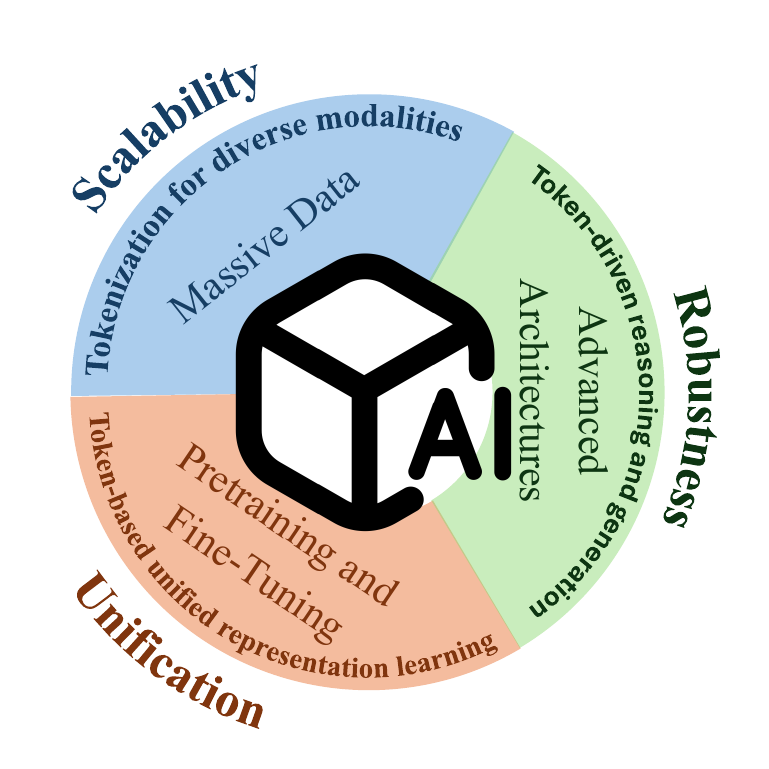}
    \caption{Characteristics of LMs and advantages of tokens}
    \label{fig:AI} 
\end{figure}
As summarized in Figure~\ref{fig:AI}, tokens offer several advantages.
\begin{itemize}
    \item \textbf{Unification}: Tokens provide a standard semantic representation, allowing different modalities to be processed and understood in the same space, and transmitted either as symbol indices or as embedding vectors.
    \item \textbf{Robustness}: Tokens degrade under channel errors in ways that LMs can compensate for, either by contextual recovery of corrupted indices or by tolerating small perturbations of embeddings.
    \item \textbf{Scalability}: Tokens can be dynamically extended with different types, either through modality-specific projectors or by appending new symbols to the LM vocabulary, to support a wide range of applications, facilitating integration with LMs in communication networks.
\end{itemize}

\subsubsection{Discrete Tokens}\label{sususec:discrete} 

Discrete tokens are symbolic representations produced by a tokenizer that maps input signals into a finite symbol set $\mathcal{V} = \{1, 2, \ldots, K\}$. For a given input $\mathbf{X}$, the discrete token sequence is denoted as
\begin{equation} 
\left\{
\begin{aligned}
    \mathbf{T}_{\mathrm{dis}} &= \mathrm{Tokenizer}(\mathbf{X}) {}= [t_1, t_2, \ldots, t_L] \in \mathcal{V}^{L} , \\
    t_i &{}= \arg\min_{k \in \mathcal{V}} \big\| \mathbf{u}_i - \mathbf{c}_k \big\|_2 , \quad \text{(quantization-based tokenizers)}
\end{aligned}
\right.
\label{eq:discrete_token}
\end{equation}
where $L$ is the sequence length. For text, tokens are obtained through
vocabulary lookup. For other modalities, a modality-specific encoder first produces continuous features $\mathbf{u}_i \in \mathbb{R}^{d_{\mathrm{c}}}$, $i = 1, \ldots, L$, each of which is mapped by the second line to the index
of its nearest codeword in a learned codebook $\mathcal{C}=\{\mathbf{c}_1,\mathbf{c}_2,\ldots,\mathbf{c}_K\}$
with $\mathbf{c}_k \in \mathbb{R}^{d_{\mathrm{c}}}$, where $d_{\mathrm{c}}$ denotes the codeword dimension. Before entering the LM, a discrete token sequence is mapped back to vectors through an embedding lookup.

Discrete tokens possess several notable characteristics. First, their symbolic nature makes each token enumerable and addressable, which supports entropy coding, token-level probability modeling and per-token retransmission. Second, although discrete tokens may be sensitive to channel perturbations, the strong prior knowledge embedded in LMs enables corrupted or erased tokens to be concealed by context-based regeneration, thus improving robustness in practical SemCom deployments. Finally, discrete representations are natively carried by digital bit streams, and are therefore directly compatible with existing digital communication infrastructure.

A representative line of work on discrete visual tokenization is based on vector quantization (VQ). VQ-VAE~\cite{oordNeuralDiscreteRepresentation2017} and VQGAN~\cite{esserTamingTransformersHighResolution2021} encode images into discrete codebook vectors, where VQGAN further incorporates adversarial training to enhance image clarity and detail, laying the foundation for token-based image generation with transformers.

Building upon VQGAN, recent works have further advanced discrete visual tokenization. MaskGIT~\cite{changMaskGITMaskedGenerative2022a} introduced a bidirectional transformer that leverages masked visual token modeling (MVTM) for training and iterative parallel decoding for inference, significantly accelerating image generation compared to autoregressive methods. TiTok~\cite{yuImageWorth322024a} proposed a 1D tokenization approach that represents images with as few as 32 tokens, substantially reducing the latent space dimensionality and improving both encoding and generation efficiency.

More recently, Beyer et al.~\cite{beyerHighlyCompressedTokenizer2025a} demonstrated that highly compressed 1D tokenizers possess surprising generative and editing capabilities even without training a separate generative model. Their results showed that direct token manipulation and simple test-time optimization enable tasks such as inpainting and text-guided editing by exploiting the strong semantics encoded in the compressed latent space. These findings indicate that discrete tokens can serve as compact semantic representations, with direct implications for efficient SemCom.

\begin{figure*}[t]
\centering
\includegraphics[width=0.9\textwidth]{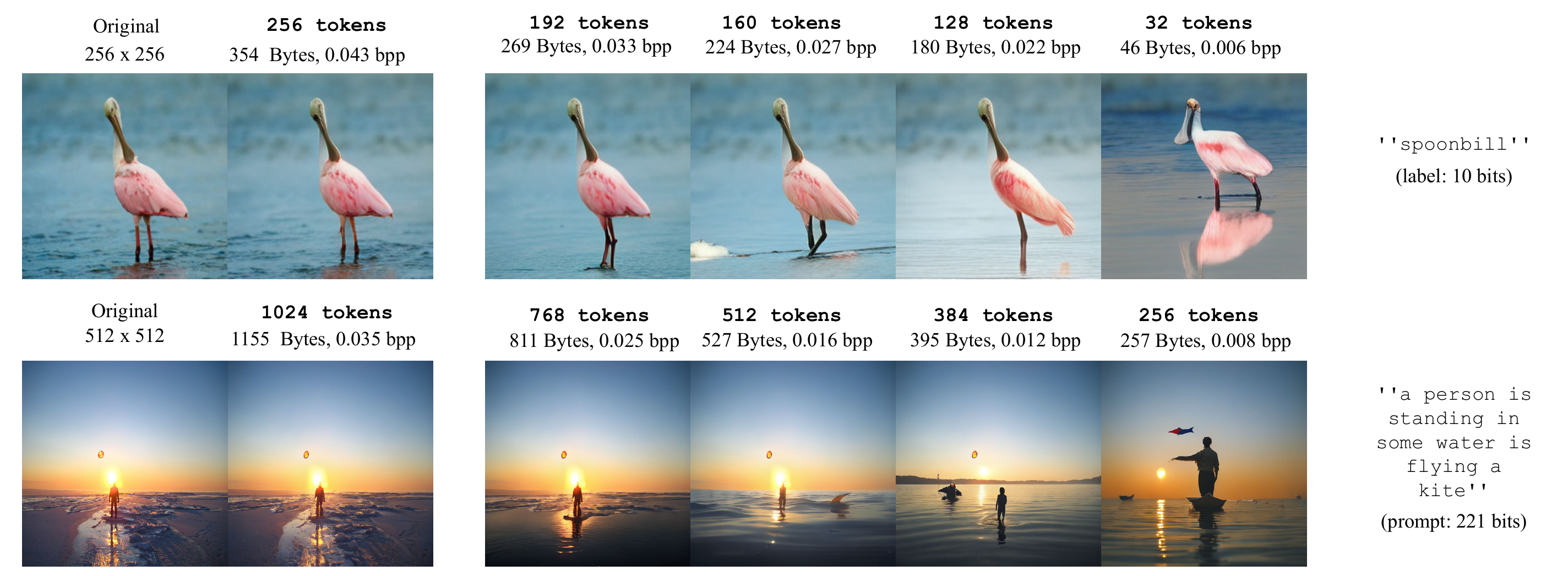}
\caption{Quality-controllable image tokenization with Ada-TokenCom~\cite{zhangAdaTokenComRateAdaptive2026}: reconstructions from a decreasing number of transmitted tokens.}
\label{fig:tokenizer_compression} 
\end{figure*}

LlamaGen~\cite{sunAutoregressiveModelBeats2024} scaled the VQGAN-style tokenizer with a substantially larger codebook (16,384 entries) and showed that plain Llama-style autoregressive transformers over such tokens can surpass diffusion models on class-conditional image generation. One-D-Piece~\cite{miwaOneDPieceImageTokenizer2025} augmented 1D tokenization with a tail-token-drop mechanism, producing variable-length token streams whose leading tokens carry the most salient semantics and thereby enabling quality-controllable compression with 1 to 256 tokens per image. As illustrated in Figure~\ref{fig:tokenizer_compression}, a discrete token sequence allows the reconstruction quality to be traded off smoothly against the number of transmitted tokens, since a receiver-side model can predict the tokens that were not sent from the transmitted prefix and a compact semantic condition.

These tokenizers further differ in the prediction order of the generative prior built on top of them, which has direct implications for transmission. AR priors over a two-dimensional token grid conventionally flatten it in raster-scan order, that is, row by row from left to right, and predict one index at a time, so that a truncated prefix does not decode to a complete image. Hierarchical alternatives instead spend several indices on the same spatial location: RQ-VAE~\cite{leeAutoregressiveImageGeneration2022} stacks residual codes at a fixed resolution, whereas visual autoregressive modeling (VAR)~\cite{tianVisualAutoregressiveModeling2024} distributes them over a coarse-to-fine pyramid of token maps and predicts one entire scale at a time. VAR uses 10 scales holding 680 code indices for a $256\times256$ image, so that generation takes 10 sequential forward passes rather than the 256 index-by-index steps of a $16\times16$ raster-ordered grid. Since each scale beyond the first refines the residual of the coarser ones, a prefix ending at a scale boundary already decodes to a complete, though coarse, full-frame reconstruction. Table~\ref{tab:tokenizers} compares representative discrete visual tokenizers.

\begin{table*}[t]
\centering
\caption{Comparison of representative discrete visual tokenizers}
\label{tab:tokenizers} 
\small
\setlength{\tabcolsep}{3pt}
\begin{tabular}{|l|c|c|c|c|c|l|}
\hline
\textbf{Tokenizer} & \textbf{Year} & \textbf{Structure} & \textbf{Codebook} & \textbf{Tokens} & \textbf{Input size} & \textbf{Generation paradigm} \\
\hline
VQ-VAE~\cite{oordNeuralDiscreteRepresentation2017} & 2017 & 2D grid & 512 & $32{\times}32$ & $128^2$ & Raster-scan AR (PixelCNN prior) \\
VQGAN~\cite{esserTamingTransformersHighResolution2021} & 2021 & 2D grid & 1,024 & $16{\times}16$ & $256^2$ & Raster-scan AR transformer \\
MaskGIT~\cite{changMaskGITMaskedGenerative2022a} & 2022 & 2D grid & 1,024 & $16{\times}16$ & $256^2$ & Masked parallel decoding \\
RQ-VAE~\cite{leeAutoregressiveImageGeneration2022} & 2022 & 2D grid, depth 4 & 16,384 & $8{\times}8{\times}4$ & $256^2$ & Raster-scan AR, depth-stacked \\
TiTok~\cite{yuImageWorth322024a} & 2024 & 1D sequence & 4,096 & 32 & $256^2$ & Masked parallel decoding \\
LlamaGen~\cite{sunAutoregressiveModelBeats2024} & 2024 & 2D grid & 16,384 & $16{\times}16$ & $256^2$ & Raster-scan AR next-token \\
VAR~\cite{tianVisualAutoregressiveModeling2024} & 2024 & 2D pyramid & 4,096 & 680 & $256^2$ & Next-scale AR (coarse-to-fine) \\
One-D-Piece~\cite{miwaOneDPieceImageTokenizer2025} & 2025 & 1D, variable length & 4,096 & 1--256 & $256^2$ & Masked parallel, variable length \\
\hline
\end{tabular}
\end{table*}

\subsubsection{Continuous Tokens} 

Continuous tokens refer to real-valued vector representations obtained by mapping multimodal inputs directly into a shared semantic embedding space, without any vocabulary lookup or quantization step. Given an input modality $\mathbf{I}$, the continuous token representation is typically obtained by projecting the features extracted by a modality-specific encoder into the token embedding space of the LM:
\begin{equation} 
    \mathbf{T}_{\mathrm{cont}} = \mathrm{Projector}\big(\mathrm{Encoder}(\mathbf{I})\big).
    \label{eq:cont_enc}
\end{equation}

Continuous tokens exhibit several advantages that make them highly suitable for multimodal SemCom. Their dense, high-dimensional nature allows rich semantic information to be preserved, enabling fine-grained representation beyond the capacity of symbolic vocabularies. Furthermore, when continuous tokens are transmitted in analog form, with each entry of the embedding vector mapped to a channel symbol, channel noise slightly displaces the received vector instead of replacing it with an unrelated token. Performance therefore degrades gracefully as the signal-to-noise ratio (SNR) decreases, rather than dropping abruptly. Continuous tokens also enable straightforward integration of heterogeneous modalities, since different projectors can map audio, vision, sensor data, or other inputs into the same semantic space without requiring modality-specific vocabularies. This property significantly enhances scalability and supports the development of unified multimodal communication frameworks.

Table~\ref{tab:token_types} contrasts the two token families from a communication perspective. Notably, discrete tokens are naturally transmitted as digital bit streams, whereas continuous tokens lend themselves to analog transmission in which the elements of the embedding vector are directly mapped to channel symbols in a JSCC fashion.

\begin{table*}[t]
\centering
\caption{Discrete versus continuous tokens from a communication perspective}
\label{tab:token_types} 
\small
\setlength{\tabcolsep}{6pt}
\begin{tabular}{|p{2.7cm}|p{6.6cm}|p{6.6cm}|}
\hline
\textbf{Aspect} & \textbf{Discrete tokens} & \textbf{Continuous tokens} \\
\hline
Representation & Indices in a finite codebook or vocabulary & Real-valued embedding vectors \\
\hline
Interpretability & Arbitrary labels; index proximity carries no semantic relation & Embedding distance correlates with semantic similarity in spaces trained or aligned for it \\
\hline
Compression & Index only; highly compact & Full vector; quantization needed for a finite rate \\
\hline
LM integration & Input by embedding lookup; output by the native softmax head & Input by a trained projector; output needs a non-categorical head \\
\hline
Transmission and robustness & Digital channel coding; residual erasures regenerated from context, avoiding retransmission & Analog JSCC-style symbol mapping; small perturbations give graceful degradation \\
\hline
\end{tabular}
\end{table*}

\section{Ubiquitous Intelligence in 6G Networks}\label{sec:Intelligence} 

\begin{figure}[tbp]
\vspace*{-3mm}
\begin{center}
    \includegraphics[width=0.45\linewidth]{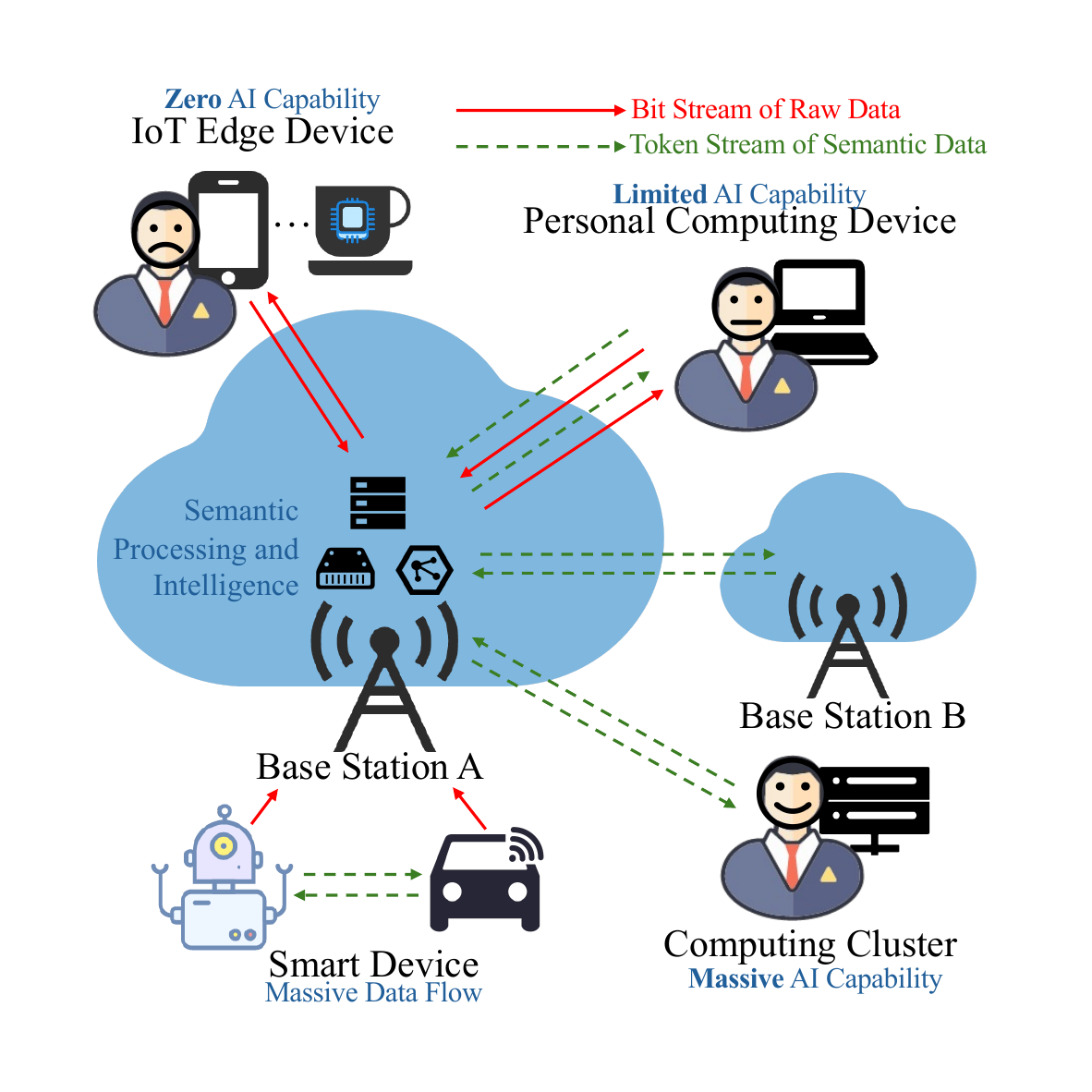}
		\end{center}
		\vspace{-8mm}
\caption{Data and computation distribution in a SemCom system}
\label{fig:intelligence} 
\end{figure}

In future wireless networks, intelligence will become a key foundation~\cite{youWhenAIMeets2025} as the Internet of Everything (IoE) evolves. By interconnecting people, objects, and data through computing-power networks, communication systems will support integrated perception, decision-making, and execution. Beyond data transport, next-generation networks will also serve as collaborative platforms for intelligent agents. As key enablers of artificial general intelligence (AGI), LMs will play a central role in shaping these intelligent communication infrastructures.

In this paradigm, everyday objects will function as intelligent nodes, such as smart cups that track hydration and chairs that monitor posture. Embedded sensors and chips will transform devices into networked terminals. As connected devices proliferate, analytics that today run in a centralized cloud will be distributed toward the network edge. For resource-limited devices, computation will increasingly migrate to communication base stations, which will process and analyze multimodal sensor data during transmission.

As shown in Figure~\ref{fig:intelligence}, devices in future networks exhibit heterogeneous AI capabilities. Some edge or IoE devices lack the ability to perform semantic processing, while others can extract partial semantic features. Only powerful nodes such as base stations or computing clusters can complete full semantic inference. Consequently, raw-data bit streams and semantic-data token streams coexist in the system. The bit streams transmitted by low-capability devices are further processed at the base stations or cloud clusters to generate or refine semantic information. This distributed processing framework enables flexible cooperation between edge and cloud, improving efficiency across heterogeneous computational resources.

Realizing this vision requires highly efficient protocols and architectures to support massive connectivity and real-time analytics. Integrating LM computing into wireless networks enables intelligent processing during transmission, which reflects the synergy between LMs and SemCom for intelligent interconnection.

\begin{table*}[b]
\centering
\caption{Example traffic characteristics and service requirements for token communications. Adapted from~\cite{3GPP2025FLS6GR}}
\label{tab:token_traffic} 
\small
\begin{tabular}{|p{2.4cm}|c|c|c|c|}
\hline
\textbf{Scenario} & \textbf{Token rate} & \textbf{Token size (bits/token)} & \textbf{Success rate} & \textbf{Delay budget} \\
\hline

Human\,$\leftrightarrow$\,Agent
& 30K$\sim$100K/s 
& 10$\sim$20 (small), $\sim$400 (large)
& 99.9\% (text), 80$\sim$99\% (others) 
& 0.1$\sim$1 s \\

\hline

Robot\,$\leftrightarrow$\,Agent
& 30K$\sim$60K/s 
& 10$\sim$20 (small), $\sim$400 (large)
& 80$\sim$99\% 
& 10$\sim$15 ms \\

\hline

Agent\,$\leftrightarrow$\,Agent 
& up to 30K/s 
& $\sim$20 bits 
& 99.9\% (text), 90$\sim$99\% (others) 
& 1$\sim$15 ms \\

\hline
\end{tabular}
\end{table*}

TokenCom has emerged as a representative traffic paradigm for future AI-native services in 6G networks, particularly driven by generative AI and intelligent agent interactions. As proposed in a company contribution to 3GPP RAN1 on evaluation assumptions for the 6G radio air interface~\cite{3GPP2025FLS6GR}, token-based services exhibit distinctive traffic patterns that differ significantly from conventional packet-oriented models.

As summarized in Table~\ref{tab:token_traffic}, token traffic is characterized by extremely high arrival rates combined with small payload sizes, posing new challenges for radio access networks in terms of control overhead, scheduling efficiency, and small-packet transmission capability. Furthermore, TokenCom exhibits heterogeneous reliability and latency requirements: text tokens typically demand success rates up to 99.9\% to preserve semantic integrity, whereas visual tokens may tolerate lower success rates of 80--99\%. These differentiated requirements highlight that future 6G systems must support token-level quality differentiation, motivating the introduction of token-oriented traffic models as a key component in 6G system design.

\section{Semantic Communication}\label{sec:Semcom} 

SemCom involves not only the transmission of information but also its comprehension and processing. In existing research, aside from LM-driven SemCom, several approaches employing lightweight models have also been proposed. Zhang et al.~\cite{zhangOptimizationImageTransmission2024} established a mapping from images to scene graphs by leveraging an object recognition model and semantic triple relationships, thereby relying on a hand-crafted extraction pipeline and a predefined semantic format. However, this approach exhibits limited flexibility, lacks attention to fine-grained details, and shows weak generalization across diverse scenarios. Furthermore, it does not incorporate an effective semantic recovery mechanism.

Consequently, LM-driven SemCom has gradually become the mainstream paradigm. By integrating computational power and intelligence into the communication process, it enables more efficient utilization of spectrum resources, enhances communication efficiency, and strengthens the network's autonomous decision-making capability. In this section, we review recent advancements in SemCom technologies along three directions: \textbf{Source-Centric Semantic Coding} focuses on semantic extraction and compression, \textbf{Channel Semantics for Physical-Layer Tasks} addresses channel modeling, estimation, feedback, and prediction, as well as channel-related signal detection and beamforming, and \textbf{Collaborative Edge-Device Intelligence} explores distributed architectures for deploying LMs across network nodes. Table~\ref{tab:semcom_comparison} compares representative studies reviewed in this section across source modalities, coding paradigms, underlying model types, and design targets. It shows that SemCom systems are shifting from conventional task-specific deep learning networks (DL) toward pre-trained LMs, and that channel-semantic works such as large wireless model (LWM)~\cite{alikhaniLargeWirelessModel2025} fall entirely outside the classical dichotomy between SSCC and JSCC, indicating that a new communication framework is needed --- a gap that the token-based paradigm in Section~\ref{sec:TokenCom} aims to fill.

\begin{table*}[t]
\centering
\caption{Overview of representative SemCom studies reviewed in Section~\ref{sec:Semcom}, by source modality, coding paradigm, model type, and design target. DL: conventional task-specific deep learning networks; LM: pre-trained large/foundation model; GAI: generative model (GAN/VAE/diffusion); Recon.: reconstruction fidelity; Task: downstream task performance}
\label{tab:semcom_comparison} 
\small
\setlength{\tabcolsep}{4.5pt}
\begin{tabular}{|l|c|l|cc|ccc|cc|}
\hline
 & & \textbf{Source} & \multicolumn{2}{c|}{\textbf{Coding}} & \multicolumn{3}{c|}{\textbf{Model type}} & \multicolumn{2}{c|}{\textbf{Target}} \\
\textbf{Reference} & \textbf{Year} & \textbf{modality} & SSCC & JSCC & DL & LM & GAI & Recon. & Task \\
\hline
DeepSC~\cite{xieDeepLearningEnabled2021} & 2021 & Text & \xmark & \cmark & \cmark & \xmark & \xmark & \cmark & \xmark \\
DeepJSCC~\cite{bourtsoulatzeDEEPJOINTSOURCECHANNEL} & 2019 & Image & \xmark & \cmark & \cmark & \xmark & \xmark & \cmark & \xmark \\
Qiao et al.~\cite{qiaoLatencyAwareGenerativeSemantic2024} & 2024 & Image & \cmark & \xmark & \xmark & \cmark & \cmark & \cmark & \xmark \\
LAM-SC~\cite{jiangLargeAIModelBased2023} & 2024 & Image & \cmark & \xmark & \cmark & \cmark & \xmark & \cmark & \xmark \\
D$^2$-JSCC~\cite{huangD2JSCCDigitalDeep2024} & 2025 & Image & \xmark & \cmark & \cmark & \xmark & \xmark & \cmark & \xmark \\
Yuan et al.~\cite{yuanGenerativeSemanticCommunication2024} & 2025 & Image & \xmark & \cmark & \cmark & \xmark & \cmark & \cmark & \cmark \\
DeepSC-S~\cite{wengSemanticCommunicationSystems2021} & 2021 & Speech & \xmark & \cmark & \cmark & \xmark & \xmark & \cmark & \xmark \\
DeepSC-ST~\cite{wengDeepLearningEnabled2023} & 2023 & Speech & \xmark & \cmark & \cmark & \xmark & \xmark & \cmark & \cmark \\
Jiang et al.~\cite{jiangWirelessSemanticCommunications2023} & 2023 & Video & \cmark & \xmark & \cmark & \xmark & \cmark & \cmark & \xmark \\
A-GSC~\cite{yangAgentDrivenGenerativeSemantic2025} & 2025 & Video & \cmark & \xmark & \cmark & \xmark & \cmark & \cmark & \xmark \\
GVSC~\cite{11112664} & 2026 & Video & \cmark & \cmark & \xmark & \cmark & \cmark & \cmark & \xmark \\
LGVSC~\cite{maLGVSCLargeModelDriven2026} & 2026 & Video & \cmark & \cmark & \xmark & \cmark & \cmark & \cmark & \cmark \\
MU-DeepSC~\cite{xieTaskOrientedMultiUserSemantic2022} & 2022 & Multi. & \xmark & \cmark & \cmark & \xmark & \xmark & \xmark & \cmark \\
SyncSC~\cite{tianSynchronousMultiModalSemantic2025} & 2025 & Multi. & \cmark & \xmark & \cmark & \cmark & \cmark & \cmark & \xmark \\
VLM-CSC~\cite{jiangVisualLanguageModelBased2025} & 2025 & Multi. & \cmark & \xmark & \cmark & \cmark & \cmark & \cmark & \xmark \\
ISCom~\cite{huangISComInterestAwareSemantic2024} & 2024 & 3D & \cmark & \xmark & \cmark & \xmark & \xmark & \cmark & \xmark \\
PCSC~\cite{liuSemanticCommunicationSystem2025} & 2025 & 3D & \xmark & \cmark & \cmark & \xmark & \xmark & \cmark & \xmark \\
LWM~\cite{alikhaniLargeWirelessModel2025} & 2024 & CSI & \xmark & \xmark & \xmark & \cmark & \xmark & \xmark & \cmark \\
\hline
\end{tabular}
\end{table*}

\subsection{Source Semantics vs. Channel Semantics} 

\begin{figure*}[t]
    \centering
    \includegraphics[width=0.9\textwidth]{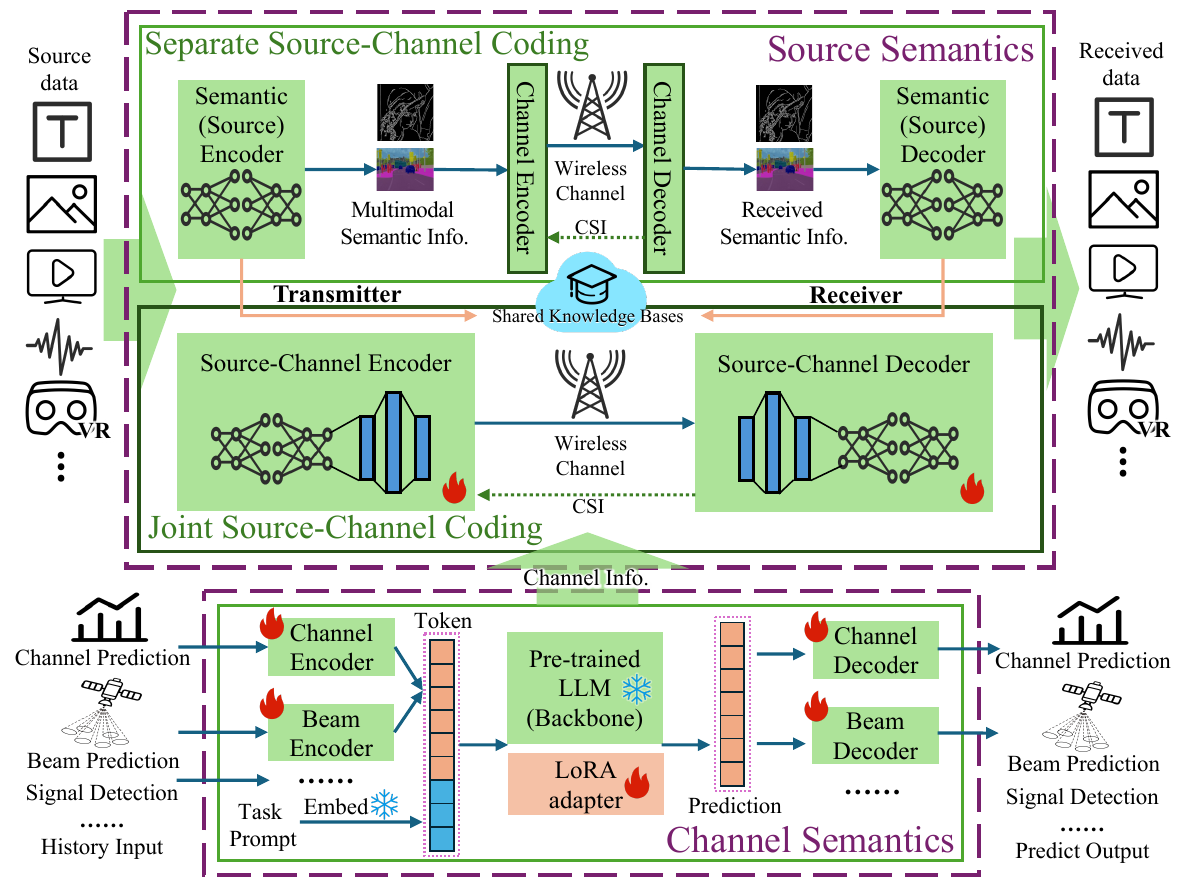}
    \caption{Basic framework of a SemCom system}
    \label{fig:SemCom} 
\end{figure*}

As illustrated in Figure~\ref{fig:SemCom}, source semantics and channel semantics respectively focus on efficient information understanding and compression during transmission, and on channel feature extraction and adaptation. Recent studies have begun to explore the application of LMs in both directions, where they play distinct functional roles.

For source semantics, LMs primarily act as semantic encoders and decoders, endowing communication systems with deeper comprehension and stronger compression capabilities. At the transmitter, a pre-trained LM encodes user information into significantly fewer bits by retaining only what the receiver's task requires and discarding redundant data, thereby reducing transmission overhead. Such semantic understanding and task-oriented compression align well with the core strengths of multimodal LMs. The resulting representations take modality-dependent forms, including textual descriptions, semantic segmentation maps or boundary graphs for images, and key frames for videos. At the receiver, a machine receiver can directly interface with downstream tasks for semantic data processing, whereas a human receiver relies on generative LMs to perform controllable, conditional generation or reconstruction based on the retained semantic features.

Research on source semantics also considers channel conditions, giving rise to separate and joint source-channel coding paradigms. In separate schemes, LMs first perform semantic encoding, and the resulting representation is then protected by a separate channel encoder. In contrast, joint schemes merge the two stages into an end-to-end design that maximizes overall system performance.

For channel semantics, the focus shifts toward exploiting the strong feature extraction and representation capabilities of LMs. By training communication-oriented LMs on large-scale channel data, the system learns compact channel features that are reused across physical-layer tasks, including channel estimation, CSI feedback and prediction, signal detection, and beamforming design. Once interpreted by the LM, channel information can be viewed as a specialized semantic representation that guides the SemCom system toward more efficient encoding and transmission strategies.

\subsection{Source Semantic Communication} 

In source semantic communication, LM-enabled compression and reconstruction are realized through concrete transmission schemes. LMs have demonstrated the ability to perform effective semantic compression even at extremely low bitrates, such as $100$~$\mu$bits per pixel (bpp) at $1024\times1024$ resolution~\cite{dotzelEXPLORINGLIMITSSEMANTIC2024}, and to reconstruct visual content in a highly controllable manner under multimodal guidance~\cite{zhanMultimodalImageSynthesis2023}. These studies provide strong evidence that LMs can serve as effective semantic encoders and decoders. The schemes reviewed below differ in how channel coding is incorporated: it is performed separately from semantic coding in SSCC, integrated with semantic coding in JSCC, or combined across the two paradigms in fusion schemes. To make this structure explicit, Table~\ref{tab:sscc_jscc} summarizes the differences between SSCC and JSCC and classifies the systems reviewed in this subsection by source modality.

\begin{table*}[t]
\centering
\caption{Contrast between the SSCC and JSCC paradigms for source semantic communication (upper part) and classification of the reviewed systems by source modality (lower part), where each system is classified by its own coding architecture rather than by the subsection in which it is cited}
\label{tab:sscc_jscc} 
\small
\setlength{\tabcolsep}{4pt}
\begin{tabular}{|>{\raggedright\arraybackslash}p{2.4cm}|p{6.95cm}|p{6.95cm}|}
\hline
\textbf{Aspect} & \textbf{SSCC: separate semantic \& channel coding} & \textbf{JSCC: jointly trained end-to-end coding} \\
\hline
Architecture & Two-stage and modular: modality-specific semantic extraction produces compact semantics, which are compressed into bit streams and protected by conventional channel coding & A single neural network maps the source directly to channel symbols~\cite{bourtsoulatzeDEEPJOINTSOURCECHANNEL}, or distinct semantic and channel codecs are trained jointly~\cite{xieDeepLearningEnabled2021} \\
\hline
Semantic extraction & Off-the-shelf pre-trained models plugged in per modality, e.g., CLIP~\cite{radford2021learningtransferablevisualmodels} (image), YOLO~\cite{redmonYouOnlyLook2016} (objects), Wav2vec and speech recognition models~\cite{schneiderWav2vecUnsupervisedPretraining2019,radfordRobustSpeechRecognition2022} (speech), and PointNet~\cite{qiPointNetDeepLearning2017} (3D point cloud) & Learned implicitly by convolutional or transformer-based encoders trained together with the channel codec, without a stand-alone extraction stage \\
\hline
Optimization & Each stage optimized independently; receiver-side generative models recover content without end-to-end training~\cite{qiaoLatencyAwareGenerativeSemantic2024} & Single end-to-end objective over source distortion, semantic fidelity, or downstream-task accuracy, see Eq.~(\ref{eq:djscc_framework}) \\
\hline
Channel behavior & Cliff effect under extreme channel conditions~\cite{bourtsoulatzeDEEPJOINTSOURCECHANNEL}; mitigated by masking-based semantic completion~\cite{tianSynchronousMultiModalSemantic2025} & Graceful degradation as channel quality varies; further enhanced by adaptive coding~\cite{zhangPredictiveAdaptiveDeep2023,wangChannelAwareDeepJoint2025} and diffusion-based denoising~\cite{guoDiffusionDrivenSemanticCommunication2024,xuSemanticPriorAided2025} \\
\hline
Digital compatibility & Native: the bit-stream interface reuses existing coding and modulation stacks & Analog symbols by default; digital variants restore constellation compatibility~\cite{tungDeepJSCCQConstellation2022,huRobustSemanticCommunications2023} or a bit-stream interface~\cite{huangD2JSCCDigitalDeep2024} \\
\hline
Interpretability and flexibility & High: explicit human-readable semantics (texts, sketches, semantic maps), plus freely composable auxiliary modules (memory, masking) & Limited: learned latent features lack explicit semantic modeling; interpretability partly regained by transmitting shared-codebook indices~\cite{zhangImprovingLearningBasedSemantic2025} \\
\hline
Main limitations & Design freedom hinders a unified architecture and global optimality; LM inference cost at both ends~\cite{renGenerativeSemanticCommunication2025} & Performance tied to the alignment between training and real channels; difficult to deploy over practical multi-hop links~\cite{bianDeepJointSourceChannel2025} \\
\hline
\multicolumn{3}{|l|}{\textbf{Representative systems by source modality}} \\
\hline
Text & --- & DeepSC~\cite{xieDeepLearningEnabled2021}, Liu et al.~\cite{liuKnowledgeDistillationBasedSemantic2024} \\
\hline
Image & Qiao et al.~\cite{qiaoLatencyAwareGenerativeSemantic2024}, Liu et al.~\cite{liuCommunicateLessSynthesize2024}, Ren et al.~\cite{renGenerativeSemanticCommunication2025} & \textit{analog:} DeepJSCC~\cite{bourtsoulatzeDEEPJOINTSOURCECHANNEL}, DeepJSCC-f~\cite{kurkaCBRJiSuanDeepJSCCfDeepJoint2020}; \textit{digital:} DeepJSCC-Q~\cite{tungDeepJSCCQConstellation2022}, D$^2$-JSCC~\cite{huangD2JSCCDigitalDeep2024}, masked VQ-VAE~\cite{huRobustSemanticCommunications2023}; \textit{hybrid:} h-DJSCC~\cite{bianDeepJointSourceChannel2025}; \textit{channel-adaptive:} PADC~\cite{zhangPredictiveAdaptiveDeep2023}, CA-DJSCC~\cite{wangChannelAwareDeepJoint2025}, ConvSC~\cite{huDeepLearningBasedSemantic2025}, DeepJSCC-MIMO~\cite{10597355}, Tang et al.~\cite{tangContrastiveLearningBasedSemantic2024}; \textit{semantic-noise robust:} Peng et al.~\cite{pengRobustImageSemantic2025}; \textit{diffusion denoising:} Guo et al.~\cite{guoDiffusionDrivenSemanticCommunication2024}, Pei et al.~\cite{peiLatentDiffusionModelEnabled2025}, Xu et al.~\cite{xuSemanticPriorAided2025} \\
\hline
Speech & --- & DeepSC-S~\cite{wengSemanticCommunicationSystems2021}, DeepSC-ST~\cite{wengDeepLearningEnabled2023} \\
\hline
Video & A-GSC~\cite{yangAgentDrivenGenerativeSemantic2025} & --- \\
\hline
3D point cloud & ISCom~\cite{huangISComInterestAwareSemantic2024} & PCSC~\cite{liuSemanticCommunicationSystem2025} \\
\hline
Multimodal & SyncSC~\cite{tianSynchronousMultiModalSemantic2025} & MU-DeepSC~\cite{xieTaskOrientedMultiUserSemantic2022} \\
\hline
& \multicolumn{2}{p{\dimexpr13.9cm+2\tabcolsep+\arrayrulewidth\relax}|}{\textbf{SSCC \& JSCC fusion designs.} Image: LAM-SC~\cite{jiangLargeAIModelBased2023} and VLM-CSC~\cite{jiangVisualLanguageModelBased2025} (segmentation- and captioning-based LM knowledge bases, each with an alternately trained channel codec), SUITS~\cite{11148624} (joint coding core with separate receiver-side restoration), and shared-codebook residual coding~\cite{zhangImprovingLearningBasedSemantic2025} (SSCC-style codebook prior with jointly trained residual coding); Video: GVSC~\cite{11112664} and LGVSC~\cite{maLGVSCLargeModelDriven2026} (textual semantics over conventional coding, visual semantics over joint coding)} \\
\hline
\end{tabular}
\end{table*}

\subsubsection{Separate Source-Channel Coding} 

Early research on source semantic communication primarily focused on SSCC methods, which is conceptually simple and can be readily integrated into existing communication system architectures, offering strong practicality and compatibility. In this paradigm, modality-specific semantic extraction schemes are first designed to obtain semantic representations from various types of data. These representations are then compressed into bit streams, after which conventional channel coding is applied to protect and transmit them. At the receiver, the received semantic information is used by a generative model to reconstruct the original content.
This design paradigm can be abstracted as a two-stage pipeline of \textbf{semantic extraction} at the transmitter and \textbf{semantic recovery} at the receiver, shown in the upper part of Figure~\ref{fig:SemCom}, mainly targeting human-oriented applications and emphasizing the visual fidelity and interpretability of the transmitted semantic information.

Different modalities generally require specifically designed \textbf{semantic extraction} schemes.
For image data, pre-trained multimodal models are commonly employed to extract semantic representations. For example, CLIP~\cite{radford2021learningtransferablevisualmodels} produces a semantic feature vector in a joint image--text embedding space, whereas generative vision-language models describe the image in natural language~\cite{renGenerativeSemanticCommunication2025}. Depending on the required level of semantic fidelity, structured semantic information such as sketch images may also be transmitted to enhance semantic preservation~\cite{qiaoLatencyAwareGenerativeSemantic2024}.

For video data, video information can be regarded as an extension of image information. Semantic features are extracted frame by frame, while LMs capture inter-frame dynamic semantics~\cite{11112664}. Object detection models such as you only look once (YOLO)~\cite{redmonYouOnlyLook2016} can be applied to detect and track semantic objects in videos, encoding video information into compact structured or textual descriptions that include object categories and spatial positions~\cite{yangAgentDrivenGenerativeSemantic2025,renGenerativeSemanticCommunication2025a}. In addition, predictive networks can be designed to infer dynamic semantic variations between adjacent frames~\cite{yangAgentDrivenGenerativeSemantic2025}. Frame-level extraction can also be made adaptive. In the large-model-driven generative video semantic communication (LGVSC) framework, Ma et al.~\cite{maLGVSCLargeModelDriven2026} computed a probability-based semantic similarity score (PSSS) from a multimodal LLM's output distribution and inserted a keyframe only where the score indicates a change of content, so that segmentation follows the video's semantics rather than a fixed interval.

For speech data, semantic extraction typically involves two aspects: content information conveyed by speech and affective information such as tone and emotion. Emotional features are often represented using traditional time-domain metrics (e.g., short-term energy and zero-crossing rate) and frequency-domain parameters (e.g., fundamental frequency)~\cite{chenSurveySemanticExtraction2025}. Deep learning-based speech embeddings, such as those produced by Wav2vec~\cite{schneiderWav2vecUnsupervisedPretraining2019}, are also widely used to obtain high-level feature vectors. The extraction of speech content information is usually accomplished through automatic speech recognition (ASR) models, which convert audio signals directly into textual representations~\cite{radfordRobustSpeechRecognition2022}.

For metaverse or mixed reality scenarios, which transmit gigabit-level data per second, models such as PointNet~\cite{qiPointNetDeepLearning2017} are employed to extract high-dimensional semantic representations from 3D point clouds. In interest-aware semantic communication (ISCom), the semantic extraction stage uses the user's predicted viewing pose to retain more points in the region of interest (ROI) and downsample the rest, thereby substantially reducing the amount of data transmitted~\cite{huangISComInterestAwareSemantic2024}.

Within the SSCC framework, the flexible design of encoders enables the semantic extraction and transmission of hybrid-modality information. For example, in the synchronous multimodal semantic communication system (SyncSC), Tian et al.~\cite{tianSynchronousMultiModalSemantic2025} used a pre-trained face reconstruction model to extract 3D morphable model (3DMM) coefficients representing facial motion, and an ASR model to extract speech content features, jointly constructing a comprehensive semantic representation for video-based voice chat scenarios.

For \textbf{semantic recovery}, SSCC methods typically rely on generative LMs to perform controllable reconstruction based on semantic information, thereby achieving visually faithful semantic transmission.
For image semantic recovery, generative models such as diffusion models, GANs, or VAEs are commonly employed~\cite{qiaoLatencyAwareGenerativeSemantic2024,xiaGenerativeAISemantic2025}. Depending on the type of semantic information available, suitable network architectures can be selected or fine-tuned to enhance reconstruction fidelity.
For video semantic recovery, the generative video semantic communication (GVSC) framework of Yin et al.~\cite{11112664} directly utilized pre-trained video generation models such as Open-Sora. LGVSC~\cite{maLGVSCLargeModelDriven2026} applied the same class of generation model segment by segment, which allows videos of arbitrary length to be reconstructed at a channel bandwidth ratio of about $6\times10^{-4}$, rather than the fixed-length clips of GVSC. The agent-driven generative semantic communication (A-GSC) framework of Yang et al.~\cite{yangAgentDrivenGenerativeSemantic2025} instead trained a diffusion model conditioned on the visual layout and the local static background, and predicted intermediate layouts to keep the reconstructed video coherent.

The SSCC paradigm is also well suited for explicitly distinguishing semantic importance. Liu et al.~\cite{liuCommunicateLessSynthesize2024} proposed an SSCC framework for intent-based semantic multicasting. Their method transmits only the user-intended semantic classes, while a pre-trained diffusion model synthesizes the non-intended regions from a compressed semantic map. This approach minimizes total latency under joint communication and computation constraints.

Furthermore, to cope with continuously evolving semantic demands and dynamic channel environments, the vision-language model-based cross-modal semantic communication (VLM-CSC) system~\cite{jiangVisualLanguageModelBased2025} introduced a memory-assisted encoder and decoder (MED) and a noise attention module (NAM). The MED employs short-term memory (STM) and long-term memory (LTM) mechanisms to enable online learning of new knowledge while effectively recalling historical data distributions, thereby mitigating the catastrophic forgetting problem. Meanwhile, the NAM adaptively adjusts the weighting between semantic and channel encoding based on the SNR feedback, which enhances semantic fidelity under high SNR conditions and improves noise robustness under low SNR conditions, thereby ensuring robust semantic transmission across diverse channel environments.

Although SSCC methods are conceptually simple and easy to implement, they still face several significant challenges:
\begin{itemize}
    \item \textbf{First}, the introduction of LMs can lead to asynchronous transmission across modalities. This issue can be effectively mitigated by adding timestamps to semantic data~\cite{tianSynchronousMultiModalSemantic2025}, ensuring temporal alignment among heterogeneous modalities.
    \item \textbf{Second}, the fidelity of semantic transmission lacks a well-established forward error correction mechanism, making it difficult to handle packet loss at the semantic level. To address this, masking-based methods such as masked autoencoders~\cite{tianSynchronousMultiModalSemantic2025} and masked VQ-VAEs~\cite{huRobustSemanticCommunications2023} have been introduced. These approaches simulate packet loss during training, enabling the decoder to infer and complete missing semantic information from contextual cues, thus eliminating the need for retransmission. Alternatively, pre-trained LMs such as BERT can be directly utilized for semantic-level completion based on their strong contextual understanding~\cite{tianSynchronousMultiModalSemantic2025}.
    \item \textbf{Third}, the use of pre-trained LMs and generative networks at both the encoding and decoding ends introduces high computational complexity, resulting in non-negligible latency overhead. While edge-device collaborative frameworks have been proposed to optimize the trade-off between local computing and transmission offloading~\cite{renGenerativeSemanticCommunication2025}, we reserve the detailed discussion of such architectural optimizations for Section~\ref{subsec:Collaboration}.
\end{itemize}

Due to its modular nature, the SSCC approach allows for flexible selection of various semantic encoding models and the integration of diverse auxiliary modules (e.g., memory modules, masking modules) to enhance overall system performance. However, this high degree of design freedom also hinders the development of a unified SemCom architecture, making it difficult to achieve global performance optimality. Moreover, since source and channel coding are optimized independently, SSCC schemes still suffer from the well-known cliff effect under extreme channel conditions~\cite{bourtsoulatzeDEEPJOINTSOURCECHANNEL}.

\subsubsection{Joint Source-Channel Coding} 

To address the limitations of conventional SSCC, which suffers from the cliff effect and inefficiency in low SNR or limited bandwidth regimes, deep joint source-channel coding (DJSCC) has emerged as a promising alternative. DJSCC leverages deep neural networks (DNNs), employing an autoencoder-based architecture, to directly map source signals, such as image pixel values, to the complex-valued channel input signals, thereby bypassing the explicit quantization and error correction steps of traditional digital systems. Within this paradigm, two design styles coexist: some works learn a single encoder that maps the source directly to channel symbols, whereas others retain distinct semantic and channel encoders but train them jointly. This joint design not only outperforms SSCC under challenging channel conditions and limited bandwidth but also exhibits graceful performance degradation with varying SNR, avoiding the abrupt quality drop typical of digital schemes.
The DJSCC framework can be mathematically formulated as follows~\cite{bourtsoulatzeDEEPJOINTSOURCECHANNEL}:
\begin{equation} 
\begin{aligned}
    \mathbf{z} &= f_{\boldsymbol{\theta}}(\mathbf{x}) , \\
    \hat{\mathbf{z}} &= \eta(\mathbf{z}) = \mathbf{H} \mathbf{z} + \mathbf{n} , \\
    \hat{\mathbf{x}} &= g_{\boldsymbol{\phi}}(\hat{\mathbf{z}}) , \\
    (\boldsymbol{\theta}^*, \boldsymbol{\phi}^*) &= \arg\min_{\boldsymbol{\theta}, \boldsymbol{\phi}} \mathbb{E}_{\mathbf{x}, \hat{\mathbf{x}}} \left[ d(\mathbf{x}, \hat{\mathbf{x}}) \right] ,
\end{aligned}
\label{eq:djscc_framework}
\end{equation}
where $\mathbf{x} \in \mathbb{R}^n$ is the source signal vector, with an image as a representative example;
$\mathbf{z} \in \mathbb{C}^k$ denotes the channel input symbols;
$f_{\boldsymbol{\theta}}$ and $g_{\boldsymbol{\phi}}$ are the encoder and decoder DNNs, parameterized by $\boldsymbol{\theta}$ and $\boldsymbol{\phi}$, respectively;
$\eta(\cdot)$ represents the channel transformation modeled as a non-trainable layer, with $\mathbf{H}$ as the channel gain and $\mathbf{n}$ as the additive noise;
$\hat{\mathbf{z}}$ and $\hat{\mathbf{x}}$ are the channel-impaired received signal and the reconstructed source, respectively;
$d(\cdot, \cdot)$ is a distortion metric, for example, the mean squared error (MSE), and the expectation $\mathbb{E}[\cdot]$ is taken over the source and channel distributions.

While foundational DJSCC works established the principle of using an autoencoder to jointly optimize source and channel coding in an end-to-end manner, primarily focusing on accurate source signal reconstruction (e.g., pixel-level fidelity for images) to replace the traditional bit and symbol error rates, deep learning-enabled semantic communication (DeepSC)~\cite{xieDeepLearningEnabled2021} advanced this concept by shifting the objective from source- and symbol-level fidelity to semantic fidelity. Specifically, DeepSC is proposed for text transmission, instantiating the DJSCC architecture by integrating a transformer-based network which performs joint semantic-channel coding. Its transceiver is composed of a semantic encoder, a channel encoder, a channel decoder, and a semantic decoder, which explicitly aims to extract and transmit the meaning of sentences. The receiver is optimized through a composite loss function that minimizes semantic errors (e.g., via cross-entropy) while simultaneously maximizing system capacity (e.g., via mutual information estimation). Consequently, DeepSC extended conventional DJSCC by both integrating these functions and operating in the semantic domain, thereby achieving improved robustness, particularly in low-SNR regimes, and establishing a new performance benchmark through semantics-oriented metrics such as sentence similarity. Thus, DeepSC can be regarded as an early realization of a semantics-driven DJSCC system that links traditional joint coding techniques with the emerging vision of SemCom.

Beyond single-shot transmission, the introduction of DJSCC with feedback, such as DeepJSCC-f~\cite{kurkaCBRJiSuanDeepJSCCfDeepJoint2020}, enables multi-layer progressive refinement or successive refinement by training additional layers and exploiting channel feedback across multiple transmissions. This yields significant gains in reconstruction quality for fixed-length transmission and supports the flexibility of variable-length transmission.

While conventional DJSCC often relies on analog transmission, recent advancements have bridged the gap with digital systems. Tung et al.~\cite{tungDeepJSCCQConstellation2022} first proposed DeepJSCC-Q (Q represents quantization), which constrains channel inputs to finite digital constellations while preserving the graceful degradation property of analog DJSCC, making it more practical for deployment in existing hardware. Building upon this, Huang et al.~\cite{huangD2JSCCDigitalDeep2024} further advanced the field with the digital deep joint source-channel coding (D$^2$-JSCC) framework, which employs deep source coding with an adaptive prior model to encode semantic features according to their distributions, followed by channel coding to protect encoded features. By jointly optimizing source and channel rates, D$^2$-JSCC achieves superior end-to-end distortion performance compared to both separation-based approaches and analog JSCC, while mitigating the cliff effect~\cite{huangD2JSCCDigitalDeep2024}. Furthermore, Bian et al.~\cite{bianDeepJointSourceChannel2025} introduced the hybrid DJSCC (h-DJSCC) framework, designed for effective image transmission across mobile multi-hop networks, addressing the challenges at the intersection of analog and digital methods. These digital and constellation-constrained DJSCC approaches enable practical deployment in existing digital communication infrastructure while preserving the benefits of joint source-channel optimization.

Beyond text and image transmission, the semantic DJSCC paradigm has been extended across different source modalities. In the field of speech transmission, DeepSC-S~\cite{wengSemanticCommunicationSystems2021} pioneered an end-to-end SemCom system for high-quality speech signal reconstruction. By comparing with a benchmark system using traditional SSCC, it demonstrated the significant advantage of joint coding under low-SNR conditions. As an evolution of this task-oriented approach, DeepSC-ST (DeepSC for speech transmission)~\cite{wengDeepLearningEnabled2023} shifted entirely towards semantic content understanding. Its DJSCC framework extracts semantically relevant features from the speech spectrum using a convolutional neural network (CNN) and a recurrent neural network (RNN) network, and is jointly trained with the channel codec in an end-to-end manner. The optimization objective focuses directly on improving text recognition accuracy rather than waveform fidelity. At the receiver, the system flexibly generates high-quality speech based on robustly decoded text, utilizing a pre-trained speech synthesis model combined with speaker voiceprint information. The text information can also be directly used for tasks such as speech transcription. In point cloud transmission, Liu et al.~\cite{liuSemanticCommunicationSystem2025} proposed the point cloud-based SemCom system (PCSC) based on DJSCC. Addressing the large amount of redundant point cloud data, they designed a rate-adaptive module that discards semantically insignificant information, thus achieving controllable coding rates without requiring extensive network fine-tuning.

DJSCC has developed two core advantages, elaborated below:
\begin{itemize}
    \item \textbf{Enhanced transmission robustness}, achieved through three complementary strategies, namely, adaptive coding, improved denoising, and robustness-oriented training.
    \item \textbf{Task-oriented optimization performance}, which directly optimizes the communication link for the downstream task rather than for source reconstruction quality.
\end{itemize}

Regarding the first advantage, the first strategy for enhancing robustness introduces adaptive coding methods. Zhang et al.~\cite{zhangPredictiveAdaptiveDeep2023} proposed the predictive and adaptive deep coding (PADC) framework, which uses a variable-length encoder (DeepJSCC-V) and semantic code masking to enable flexible compression ratio. They also introduced a lightweight network to predict the peak signal-to-noise ratio (PSNR) of the reconstruction under specific settings. Based on this, PADC achieves instance-level optimization, dynamically allocating the minimum rate that satisfies quality requirements for each image, significantly enhancing bandwidth efficiency in wireless image transmission while maintaining quality. Similarly, channel-aware DJSCC (CA-DJSCC)~\cite{wangChannelAwareDeepJoint2025} integrates implicit CSI with a semantic importance vector for pruning semantic information, achieving semantic transmission with channel-adaptive dimensions. To further address complex channel environments, Hu et al.~\cite{huDeepLearningBasedSemantic2025} proposed ConvSC, a CNN-based SemCom scheme, incorporating feature-channel attention and spatial attention conditioned on channel state information to weight the relative importance of feature channels and of positions within each channel, thereby better preserving critical semantic content under varying channel conditions. Peng et al.~\cite{pengRobustImageSemantic2025} designed a multi-scale ViT that simultaneously captures fine-grained and global semantic features, effectively mitigating the impact of adversarial perturbations on semantic impairment. Wu et al.~\cite{10597355} extended adaptive coding to multiple-input multiple-output (MIMO) channels with DeepJSCC-MIMO, whose ViT self-attention learns the feature mapping and power allocation jointly from the source image and the channel state, so that one model is robust to channel estimation errors and covers different channel conditions and antenna configurations without retraining.

The second strategy for enhancing robustness employs improved denoising methods. Pei et al.~\cite{peiLatentDiffusionModelEnabled2025} used adversarial convex optimization to achieve robust GAN inversion for handling erroneous input signals. With the rise of diffusion models, their denoising capabilities have attracted considerable attention. A diffusion-based end-to-end framework was proposed by Guo et al.~\cite{guoDiffusionDrivenSemanticCommunication2024}, which models wireless signal transmission as the forward process of a diffusion model, effectively mitigating channel noise through the reverse process. Semantic ambiguities arising from out-of-domain data are addressed using lightweight latent-space adapters trained through adversarial one-shot or few-shot learning~\cite{peiLatentDiffusionModelEnabled2025}. Xu et al.~\cite{xuSemanticPriorAided2025} further proposed semantic prior-aided channel-adaptive equalization and denoising SemCom (SP-EDNSC), which treats joint equalization and denoising as an inverse problem. Through iterative latent diffusion posterior sampling, the method combines the likelihood information of the received signal with learned semantic priors to jointly mitigate fading and noise effects.

The third strategy improves robustness through the training procedure itself. Tang et al.~\cite{tangContrastiveLearningBasedSemantic2024} proposed a contrastive learning-based approach where channel noise is treated as a form of data augmentation, significantly improving semantic consistency under corruption. This method achieves notable accuracy improvements even at low bandwidth compression ratios. In a multi-user setting, Liu et al.~\cite{liuKnowledgeDistillationBasedSemantic2024} distilled student models trained without interference samples and applied them to unseen multi-user interference, improving generalization and robustness over non-distilled baselines, while also limiting the performance loss incurred when the semantic and channel encoder-decoders are compressed.

The second advantage, the task-oriented end-to-end optimization of DJSCC, enables the entire communication link to be directly optimized for downstream tasks rather than merely focusing on source reconstruction quality.
Xie et al.~\cite{xieTaskOrientedMultiUserSemantic2022} proposed MU-DeepSC, an end-to-end transmission framework designed for visual question answering (VQA) tasks. By skipping pixel-level reconstruction and directly optimizing question-answer accuracy, the model achieved a significant reduction in bandwidth requirements while maintaining comparable accuracy.
This concept has also been extended to more complex scenarios. CA-DJSCC~\cite{wangChannelAwareDeepJoint2025} jointly optimizes image reconstruction and classification tasks, demonstrating strong performance in multi-task transmission.

\subsubsection{Fusion of Separate and Joint Coding} 

Essentially, DJSCC is a data-driven approach that achieves high compression efficiency and ensures task performance through targeted joint training. However, it faces several inherent limitations. First, DJSCC generally lacks explicit modeling of semantic information, making it difficult to interpret the semantic features involved in transmission. This lack of interpretability makes it difficult to explicitly determine what semantic information should be extracted and when it should be transmitted, thereby limiting task- and context-aware transmission decisions. Moreover, DJSCC's strong performance is often confined to simulated environments and heavily depends on the alignment between real and training channel conditions. These limitations restrict its deployment in practical scenarios involving multi-hop forwarding.
In contrast, SSCC methods offer greater flexibility and can integrate external knowledge bases to enhance overall system capability~\cite{yangAgentDrivenGenerativeSemantic2025}. Consequently, many studies combine the two approaches, in which SSCC contributes to compression efficiency while joint training reinforces transmission robustness.
Jiang et al.~\cite{jiangLargeAIModelBased2023} embodied this hybrid concept in large AI model-based semantic communication (LAM-SC), where LMs serve as knowledge bases for semantic encoding, while a learned channel codec is co-adapted with the semantic codec through alternating training rather than a single end-to-end objective, thus combining both advantages. In underwater image transmission, the semantic underwater image transmission system (SUITS) embeds an unsupervised color-correction module into the DeepJSCC decoder and appends a pre-trained super-resolution network at the receiver, improving the visual quality of the restored image, particularly at low SNR~\cite{11148624}.
Zhang et al.~\cite{zhangImprovingLearningBasedSemantic2025} further introduced a shared codebook mechanism from SSCC, enabling DJSCC to transmit only residual semantic information relative to codebook indices, thereby improving transmission efficiency while retaining the interpretability of an explicit codebook.

\subsection{Channel Semantics for Physical-Layer Tasks}\label{subsec:chansem} 

Unlike source semantics, channel semantics originated from specialist model-based approaches addressing tasks such as uplink random access~\cite{8961111,9941253}, CSI feedback~\cite{9053850,9954153}, and precoding~\cite{10680080}. Studies mainly focus on leveraging LMs for efficient channel modeling, estimation, feedback, and prediction, as well as channel-related signal detection and beamforming, exploring how to migrate specialist model functionalities into a unified LM framework. The ultimate goal is to enable a single model to semantically coordinate and solve multiple physical-layer transmission tasks. Building on the multimodal LM architectures introduced in Section~\ref{sususec:llm}, existing works generally follow two directions: (1) \textbf{fine-tuning} pre-trained LLMs for wireless-specific tasks, and (2) \textbf{pretraining} wireless foundation models on large-scale wireless datasets. Table~\ref{tab:channel_lm} summarizes the reviewed works along these two directions. The first inherits a web-text corpus and meets wireless data only at the fine-tuning stage, so the backbone and the fine-tuning method carry the design. The second pretrains on channel data and needs little or no fine-tuning.

\begin{table*}[!t]
\centering
\caption{Comparison of LM-based approaches to channel semantics for physical-layer tasks. For the upper block, the pretraining columns describe the corpus of the backbone LLM rather than data reported by the cited work. n.d.: not disclosed in the cited work; ``---'': stage not used. LoRA: low-rank adaptation; UMa: urban macro; ISAC: integrated sensing and communication; LoS/NLoS: line-of-sight/non-line-of-sight}
\label{tab:channel_lm} 
\scriptsize
\setlength{\tabcolsep}{2pt}
\begin{tabular}{|>{\raggedright\arraybackslash}p{2.3cm}|>{\raggedright\arraybackslash}p{1.93cm}|>{\centering\arraybackslash}p{1.1cm}|>{\raggedright\arraybackslash}p{2.15cm}|>{\centering\arraybackslash}p{1.15cm}|>{\raggedright\arraybackslash}p{1.83cm}|>{\raggedright\arraybackslash}p{2.08cm}|>{\raggedright\arraybackslash}p{3.65cm}|}
\hline
\multirow{2}{=}{\textbf{Name}} & \multirow{2}{=}{\textbf{Backbone}} & \multirow{2}{=}{\textbf{Params}} & \multicolumn{2}{c|}{\textbf{Pre-training}} & \multicolumn{2}{c|}{\textbf{Fine-tuning}} & \multirow{2}{=}{\textbf{Downstream tasks}} \\
\cline{4-7}
 & & & \textbf{Datasets} & \textbf{Size} & \textbf{Datasets} & \textbf{Method} & \\
\hline
\multicolumn{8}{|l|}{\textit{Direction 1: fine-tuning an LLM pre-trained on text}} \\
\hline
ChannelGPT \cite{yuChannelGPTLargeModel2025} & GPT-2 & 124M & WebText & 40~GB & Multimodal environment & Few-parameter tuning & Channel generation and prediction \\
\hline
Sheng et al.~\cite{shengBeamPredictionBased2025} & GPT-2 & n.d. & WebText & 40~GB & DeepMIMO, 73.7K & Input/output modules only & mmWave beam prediction \\
\hline
Zheng et al.~\cite{zhengLargeLanguageModel2025} & LLaMA2 & 7B & Public online data & 2T tokens & 3GPP UMa, 50K/task & LoRA & Precoding, signal detection, channel prediction \\
\hline
\multicolumn{8}{|l|}{\textit{Direction 2: pretraining a foundation model on wireless data}} \\
\hline
Sheng et al.~\cite{shengWirelessFoundationModel2025} & Causal, 16 layers & n.d. & QuaDRiGa CSI, ISAC, traffic & $>$19M samples & --- & --- & Channel, angle, traffic prediction \\
\hline
LWM~\cite{alikhaniLargeWirelessModel2025} & Encoder-only, 12 layers & 600K & DeepMIMO & 820K samples & Task-specific data & Fine-tune last layers & Beam prediction, LoS/NLoS classification \\
\hline
\end{tabular}
\end{table*}

Similar to multimodal LLM fine-tuning, some studies use adapters to map wireless data into the LLM embedding space. Sheng et al.~\cite{shengBeamPredictionBased2025} modeled beam prediction as a time-series forecasting problem, segmenting historical information and extracting features before mapping them to word embeddings. They fine-tuned a GPT-2 model using the Prompt-as-Prefix technique, enabling it to perform single-task channel modeling. For multi-task demands, ChannelGPT~\cite{yuChannelGPTLargeModel2025}, also based on GPT-2, introduced modality-specific embeddings to map multimodal inputs, such as channel data (e.g., CSI) and environmental data (e.g., images, point clouds, and positions), into unified feature representations. Cross-modal fine-tuning of the pre-trained GPT-2 with multi-task output heads enabled ChannelGPT to support diverse tasks, including channel parameter generation, channel mapping, and wireless knowledge reasoning, building general-purpose capabilities toward a real-world channel foundation model. Similarly, Zheng et al.~\cite{zhengLargeLanguageModel2025} developed multiple encoding modules and applied instruction fine-tuning to a LLaMA-based model, achieving unified multi-task performance across channel prediction, multi-user precoding, and signal detection.

However, fine-tuning general-purpose LLMs introduces inefficiency, as these models retain vast linguistic knowledge irrelevant to wireless communication, potentially reducing computational efficiency and task performance. Chen et al.~\cite{chenTowardsWirelessNative2025} attributed this mismatch to fundamental differences between language and wireless intelligence: wireless models must learn electromagnetic mechanisms from raw, structurally ordered observations while satisfying real-time and signaling constraints, making mechanism-oriented reasoning more important than broad linguistic memorization. Motivated by this mismatch, Sheng et al.~\cite{shengWirelessFoundationModel2025} later proposed a decoder-only causal transformer backbone with granularity encoding to distinguish data sampled at different intervals, and a univariate decomposition technique to handle heterogeneous multivariate time series uniformly. This design demonstrated strong zero-shot generalization on unseen tasks, marking a step toward universal channel semantic modeling. Alternatively, Alikhani et al.~\cite{alikhaniLargeWirelessModel2025} introduced LWM, a task-agnostic encoder-based transformer pre-trained via self-supervised masked channel modeling. The resulting classification (CLS) token provides a compact yet semantically rich channel representation for downstream tasks.

\subsection{Collaborative Edge-Device Intelligence}\label{subsec:Collaboration} 

The deployment of LMs in SemCom systems faces a fundamental dilemma: edge devices often lack the memory and computing capacity to host full-scale models (e.g., $7$B+ parameters), while centralized cloud inference incurs high transmission latency and privacy concerns~\cite{quMobileEdgeIntelligence2025}. Consequently, edge-device collaboration has emerged as a critical architectural paradigm, enabling distributed semantic inference through the synergy of communication and computation resources.

A primary strategy involves dynamically partitioning semantic tasks across devices, edge servers, and the cloud.
Ren et al.~\cite{renGenerativeSemanticCommunication2025} proposed an edge-device collaborative SSCC framework that optimizes the trade-off between local prompt generation and remote inference. By employing a swap/leaving/joining (SLJ)-based matching algorithm, the system dynamically offloads computation tasks based on real-time channel states and device capabilities, significantly reducing end-to-end latency.
To further support AI-native services, Chen et al.~\cite{chenNetGPTNativeAINetwork2024} introduced NetGPT, a cloud-edge architecture in which the edge LM augments prompts with location-specific information, while the cloud LM generates personalized responses. The authors further suggest that the intent inference capability of the edge LM could provide a common foundation for intelligent network management and orchestration.

Beyond simple offloading, splitting the LM itself across multiple nodes allows for handling larger models.
For resource-constrained IoE scenarios, Zhang et al.~\cite{zhang2505ZuHuiCommunicationEfficientDistributedOnDevice2025} utilized tensor parallelism to split token inference across multiple edge devices. By exploiting the superposition property of wireless channels for efficient over-the-air aggregation, this approach significantly reduces the computational load on individual devices.
Furthermore, addressing the storage challenges of LMs, Chen et al.~\cite{chenSlimCachingEdgeCaching2025} proposed ``SlimCaching,'' a strategy specifically designed for MoE models. By caching only the most relevant experts at the network edge, it enables efficient distributed inference without requiring full model replication.

Collaboration becomes even more challenging in high-mobility scenarios, such as non-terrestrial networks. Chen et al.~\cite{chenSpacegroundFluidAI2025} proposed the ``space-ground fluid AI'' framework for 6G, which utilizes the predictable mobility of satellites to facilitate fluid task and model migration. This ensures continuous semantic service provisioning despite the intermittent connectivity and high dynamics of satellite-ground links.

To mitigate the inference latency of autoregressive generation, collaborative speculative decoding has been adapted for wireless SemCom scenarios, where a device-side small language model (SLM) generates draft tokens that are verified in parallel by an edge LLM~\cite{leviathanFastInferenceTransformers2023,zhengCommunicationEfficientCollaborativeLLM2025,ohUncertaintyAwareHybridInference2025}. Here the focus is on reducing SemCom end-to-end latency through collaborative deployment. The analysis of token transmission as a native cost of LM inference itself is deferred to Section~\ref{sec:TokenCom}.

\subsection{Task-Oriented and Scenario-Specific Communication} 

Task-oriented and scenario-specific communication is an important direction in AI-enabled SemCom. Instead of maximizing overall semantic fidelity, these systems are designed to meet the requirements of specific downstream tasks or application scenarios. By integrating AI intelligence into scenario-aware decision-making, such systems achieve greater efficiency in specific applications. Yuan et al.~\cite{yuanGenerativeSemanticCommunication2024} designed a task knowledge bases (KB) that map natural-language task requirements onto predefined task instructions in the form of discrete tokens by semantic similarity, using them to guide task-relevant feature selection, together with a unified JSCC encoder and task-specific decoders for image reconstruction and segmentation.
Jiang et al.~\cite{jiangWirelessSemanticCommunications2023} proposed a semantic video conferencing system optimized for scenes with minimal background and motion changes. By transmitting only facial keypoints and reconstructing frames via generative models at the receiver, the system achieved substantial bandwidth savings. It also replaced traditional cyclic redundancy check (CRC) with a semantic error detection mechanism based on keypoint deviation and fluency metrics, demonstrating scenario-aware and user-centric intelligent communication design. Similarly, in traffic monitoring videos, only dynamic vehicle semantics need to be transmitted while static background information can be omitted~\cite{yangSemanticChangeDriven2023}. In metaverse applications, Huang et al.~\cite{huangISComInterestAwareSemantic2024} proposed ISCom, which combines two-stage ROI selection with a lightweight point-cloud video codec and a DRL-based scheduler. The ROI selector uses predicted user trajectories, view-frustum information, and saliency features to reduce the amount of transmitted data, while the scheduler adapts the codec configuration to network conditions and device capabilities. Scenario-specific requirements also extend to communication security. He et al.~\cite{heMixtureofExpertsEnabledTrustworthy2024} proposed an MoE-based SemCom system in which a gating network selects specialized semantic codec experts according to user-defined security requirements, enabling the system to defend against multiple types of attacks.

Under the concept of semantic importance, multiple-access technologies have also been revisited. Unlike traditional schemes that distinguish users by physical resources, Zhang et al.~\cite{zhangModelDivisionMultiple2023} extracted high-dimensional semantic features with a model-based AI approach and constructed a model information space for source and channel features, enabling model division multiple access (MDMA). By identifying shared and personalized information among multiuser semantics, MDMA allows shared semantic content to be transmitted over a common physical channel, substantially improving bandwidth efficiency.

Overall, LM-driven SemCom systems are evolving toward greater intelligence and autonomy. However, diverse intelligent design paradigms remain fragmented and difficult to unify within communication frameworks. This fragmentation, together with the emergence of LM-native services that require direct token routing and transmission across distributed nodes, motivates the TokenCom paradigm developed in Section~\ref{sec:TokenCom}.

\section{Token Communication}\label{sec:TokenCom} 

\subsection{From SemCom to TokenCom: The AI-Native Unified Interface Above the Bit Level} 

Despite significant progress in SemCom research, a fundamental challenge remains: the absence of a unified basic unit for semantic representation and transmission. Current SemCom approaches vary widely in their definitions of semantic modalities, encoding strategies, and transmission protocols, leading to fragmented solutions that are difficult to integrate or scale. This fragmentation stems from two inherent limitations: 1)~\textbf{modality heterogeneity}, in which different data types (e.g., text, image, video, and audio) require different semantic extraction and reconstruction methods, and 2)~\textbf{model diversity}, whereby various LMs with different architectures and training objectives produce incompatible semantic representations.

TokenCom emerges as a solution to this unification challenge. It is also a natural consequence of the stage AI has now reached, in which LM inference and serving themselves depend on transmitting tokens between distributed nodes. The key insight is that tokens, as the native processing units of LMs, can serve as a universal semantic substrate. Recent breakthroughs in unified multimodal learning have demonstrated this potential. Emu3~\cite{wangMultimodalLearningNexttoken2026} showed that a single autoregressive objective can achieve coherent outputs across text, images, and video without relying on complex diffusion or compositional architectures. More significantly, Unified-IO 2~\cite{luUnifiedIO2Scaling2024} demonstrated that tokenizing text, images, audio, and actions into a shared semantic space enables a single model to process and generate across all modalities under a unified multimodal denoising objective. This work supports the central premise of TokenCom: diverse modalities can be represented and transmitted through a single token space, and the model's cross-modal understanding carries over to communication scenarios where tokens serve as both the semantic carrier and the transmission unit. Overall, these results suggest that tokens can act as a unified interface for multimodal understanding and generation, a capability that extends to communication systems.

TokenCom changes the unit in which information is represented and processed. In the discrete case it does not remove bits from the link that carries them. A token is transmitted by sending its codebook index, which a fixed-length representation encodes in $\lceil \log_2 |\mathcal{V}| \rceil$ bits and entropy coding compresses further when the token distribution is non-uniform. The resulting bit stream passes through channel coding and modulation as in a conventional digital link, and the service requirements in Table~\ref{tab:token_traffic} are correspondingly stated in bits per token. At the packet level, Qiao et al.~\cite{qiaoTokenCommunicationsLarge2025} likewise described TokenCom packets as containers of 0s and 1s that carry strong contextual information for each data flow. What changes is the granularity at which importance is assigned and errors are handled, per token rather than per bit, and the coding, modulation, and power allocation of the link are then designed around that granularity. Continuous tokens take the analog route already established by DJSCC (see Table~\ref{tab:token_types}), in which embedding components are mapped to channel symbols.

From an information-theoretic perspective, the shift toward TokenCom also reflects a fundamental change in how we evaluate semantic transmission quality. Traditional compression theory focuses on the rate-distortion trade-off, where the goal is to minimize distortion for a given bitrate. Blau and
Michaeli~\cite{blauRethinkingLossyCompression2019} generalized this into a rate-distortion-perception (RDP) trade-off, showing that constraining perceptual quality to be high elevates the rate-distortion curve, so that perceptual fidelity comes at the cost of a higher rate or greater distortion. Recent works have argued that the evaluation of SemCom systems should correspondingly shift toward the RDP trade-off~\cite{qiaoTokenCommunicationsLarge2025,ohayonCompressedImageGeneration2025a}. This perspective aligns naturally with token-based representations: since tokens can capture high-level semantic meaning beyond low-level pixels, the reconstruction objective can prioritize semantic fidelity and perceptual plausibility. The receiver's LM supplies exactly this perceptual quality, generating coherent and contextually appropriate content from the received tokens, so a token-based link is naturally described by the RDP trade-off rather than by rate and distortion alone.

The transition from SemCom to TokenCom is driven by three key advantages of unification.
\begin{itemize}
    \item \textbf{Unified Modality Representation}: Tokens provide a common format for representing diverse data modalities, enabling cross-modal semantic alignment and joint processing within a single framework.
    \item \textbf{Unified Model Capability}: By using tokens as the basic unit, a single LM can handle multiple tasks, including understanding, generation, compression, and transmission, eliminating the need for task-specific architectures.
    \item \textbf{Unified Communication Protocol}: Tokens standardize the semantic representation itself, so that encoding, error correction, and resource allocation mechanisms can be shared across different scenarios and devices.
\end{itemize}

\subsection{Token as the Fundamental Unit of Communication} 

\begin{figure*}[t]
    \centering
    \includegraphics[width=0.8\textwidth]{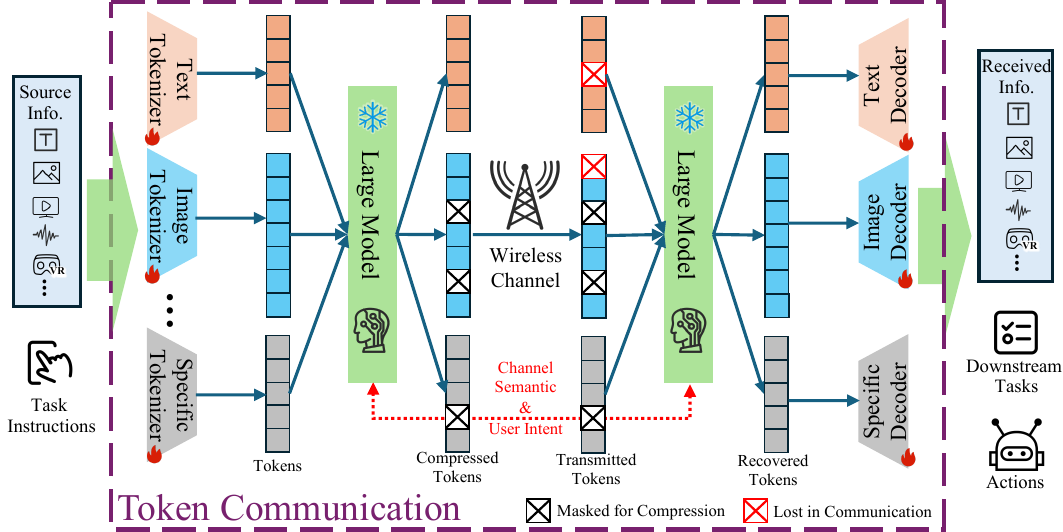}
    \caption{Token communication framework~\cite{qiaoTokenCommunicationsLarge2025}}
    \label{fig:TokenCom} 
\end{figure*}

Token naturally arises when both source and channel semantics are primarily driven by LMs, providing a unified paradigm for SemCom systems. In the communication system, the definition of the token is the \textbf{same as} that defined in LMs. The two sections differ in the role they assign to the token, not in how they define it: whereas Section~\ref{subsec:token} introduces the token as the native processing unit inside an LM, here we examine its complementary role as the fundamental transmission unit of the communication system, that is, the semantic counterpart of the bit. Tokens and bits sit at different levels of abstraction. The token is the unit at which the system reasons about meaning, while the bit is the unit at which the link transports it. Tokens and semantics differ in form, with the token serving as the unified representation that LMs across the communication system process directly. By organizing information into semantic units, it significantly enhances the system's deep understanding and processing capabilities. As the ``bit unit'' in SemCom, the token serves as the native carrier of LM information representation, meeting the intelligence needs of communication networks, such as uniform semantic data transmission and semantic-level error correction. It also enables the efficient integration of multimodal information during transmission and dynamic resource allocation based on the importance of the information, characterized by unification, robustness, and scalability.

Formally, in TokenCom, a token $t_i$ is an elementary semantic unit that, following the taxonomy of Section~\ref{subsec:token}, may take either a discrete or a continuous form. For clarity, the formulation below is presented for the discrete case, in which $t_i$ is drawn from a finite vocabulary $\mathcal{V}$, where $|\mathcal{V}|$ typically ranges from thousands to hundreds of thousands depending on the tokenizer design. In the continuous case the tokenizer is replaced by a projector-based encoder, so that the transmitted sequence consists of real-valued vectors rather than symbol indices. Unlike traditional bits that carry no inherent meaning, each token encapsulates high-level semantic information learned during LM pretraining. That is, depending on the tokenizer and representation level, a token may encode information ranging from local structural details to high-level semantics. For a source message $\mathbf{x}$ (which may be text, image, audio, video, or another modality), the tokenization process can be expressed as:
\begin{equation} 
\mathbf{x} \xrightarrow{\text{Tokenizer}} \mathbf{t} = [t_1, t_2, \ldots, t_L] \in \mathcal{V}^L ,
\end{equation}
where $L$ is the sequence length. This token sequence $\mathbf{t}$ serves as the fundamental transmission unit in TokenCom, replacing the modality- and task-specific semantic features used in conventional SemCom.

The key advantage of tokens over traditional semantic features lies in their unified, LM-native representation and semantic richness. Each token is associated with an embedding vector $\mathbf{e}_i \in \mathbb{R}^d$, obtained through an embedding layer for discrete tokens and produced directly by a projector for continuous ones, with $d$ typically ranging from 512 to 4096. This embedding captures contextual semantic information accumulated from massive pretraining data, enabling the LM to understand and generate meaningful content. Furthermore, the LM can reconstruct missing or unreliable tokens from context and channel observations rather than relying solely on bit-level error-correcting codes.

\begin{figure*}[!t]
    \centering
    \includegraphics[width=\textwidth]{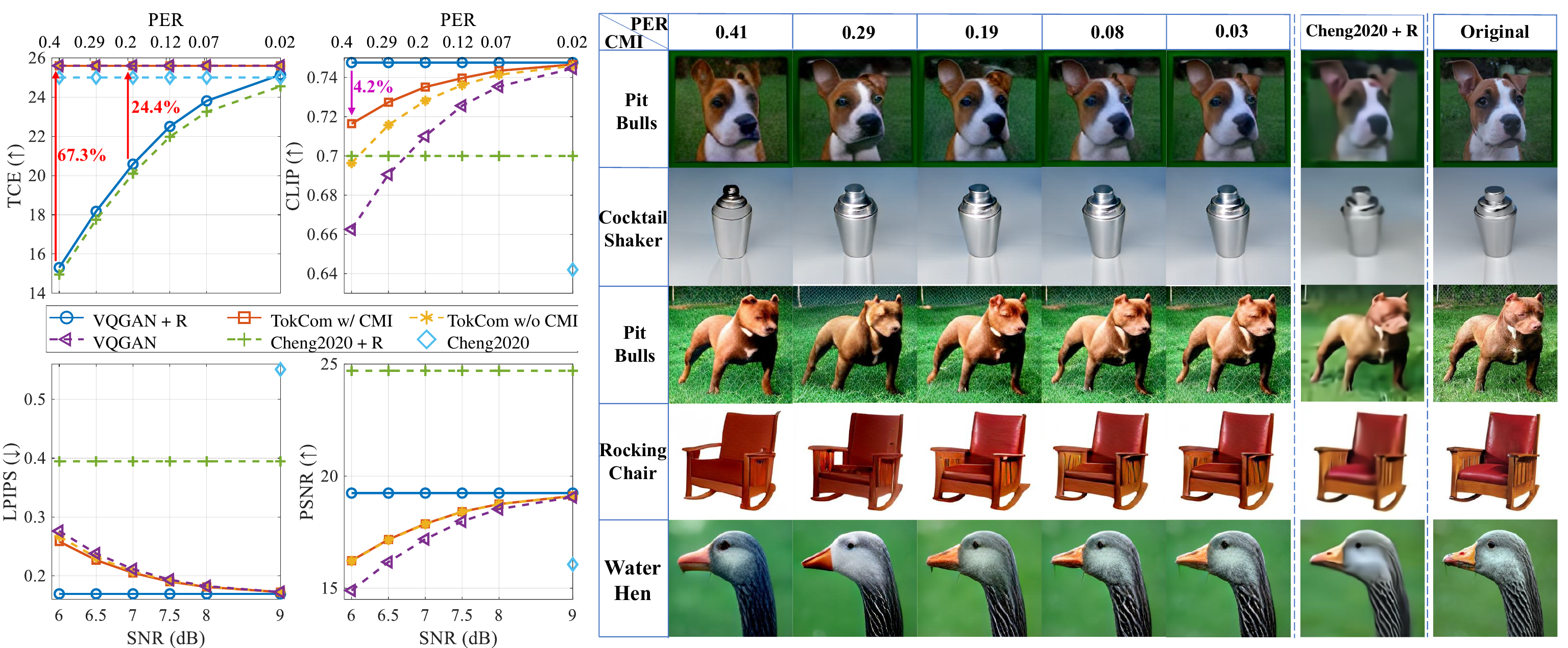}
    \caption{Simulation results for generative image TokenCom as functions of SNR and PER. TokenCom with cross-modal information (CMI) sustains high semantic similarity (CLIP), perceptual quality (learned perceptual image patch similarity, LPIPS), and pixel-level fidelity (PSNR) as the PER increases, attaining TCE gains of up to 67.3\% over conventional codebook-based baselines. The baselines are the learned image codec Cheng2020~\cite{chengLearnedImageCompression2020} and VQGAN~\cite{esserTamingTransformersHighResolution2021}, and ``+ R'' denotes a scheme in which the receiver requests retransmission of every corrupted packet. The right panel shows representative image reconstructions under decreasing PER. Reproduced from~\cite{qiaoTokenCommunicationsLarge2025}.}
    \label{fig:tokcom_perf} 
\end{figure*}

Before the concept of TokenCom was introduced, several works based on discrete codebooks had already illustrated the idea of TokenCom, using highly compressed token indices to reduce the volume of transmitted data~\cite{zhangImprovingLearningBasedSemantic2025,ohayonCompressedImageGeneration2025a}. Compared to SemCom, tokens offer advantages in semantic representation: they efficiently capture multimodal semantic information and enhance the system's semantic awareness and intelligence in a unified manner. As shown in Figure~\ref{fig:TokenCom}, each modality is converted by its own tokenizer
into a token sequence, so that heterogeneous sources enter the network in a single
common format~\cite{qiaoTokenCommunicationsLarge2025}, embodying the notion that everything can be tokenized. The performance gains of this TokenCom paradigm are also empirically quantified (see Figure~\ref{fig:tokcom_perf}): on ImageNet100 at a low source rate of 0.039 bpp (256 tokens per image, codebook size $Q=1024$, 16-point quadrature amplitude modulation (16-QAM) with rate-1/2 convolutional coding and soft Viterbi decoding), the generative masked-token receiver reduces the token error rate from 0.15 to 0.03 at 6~dB SNR and from 0.065 to 0.019 at 7~dB SNR, while keeping the CLIP semantic similarity above 0.70. At a packet error rate (PER) of 20\%, it achieves a 24.4\% gain in token communication efficiency (TCE) over the baseline. Beyond facilitating information reconstruction, tokens can also support performing downstream tasks and generating machine-level instructions during transmission, with most of the computational workload executed in the cloud. Concurrently, task-oriented token communications~\cite{zhangTaskOrientedMultimodalToken2026} have proposed and validated the feasibility of the TokenCom framework, demonstrating the potential of unified semantic representation and confirming the token's stable performance across modalities in SemCom. UniToCom~\cite{weiTokenCommunicationEra2025a} further solidified the theoretical foundation by proposing an information bottleneck-based approach to maximize semantic information retention while minimizing transmission overhead. More recently, Jiang et al.~\cite{jiangTokenComVisionLanguageModel2026} proposed TaiChi, a vision-language model-driven TokenCom framework that employs a dual visual tokenizer and a Kolmogorov-Arnold network (KAN)-based cross-modal projector, further validating the unified token paradigm for multimodal and multitask SemCom.

\subsection{Token Selection and Compression}\label{subsec:compress} 

A core advantage of TokenCom lies in its ability to perform semantic-level compression through selective transmission. Unlike traditional bit-level compression, token-based approaches can leverage the semantic understanding of LMs to determine which information needs to be transmitted. Two criteria are available for this decision. The first is \textbf{semantic importance}, which ranks tokens by how much they contribute to the downstream task or to reconstruction quality. The second is \textbf{contextual predictability}, which ranks tokens by how reliably the receiver can infer them from the surrounding context. This criterion becomes available only because the receiver holds a pre-trained model of the token sequence itself. The two criteria are not equivalent, since a token can be central to the meaning of a sentence and at the same time be almost fully determined by its neighbors.

The first criterion, \textbf{semantic importance}, is estimated at the transmitter. The LM empowered cross-modal semantic communication (LMECM-SC) system~\cite{pengLargeModelEmpowered2025} advanced this concept by leveraging a cross-modal attention mechanism to quantify the importance of each token in cross-modal interactions. The system dynamically selects the top $k\%$ of tokens most useful for downstream tasks for encoding and transmission, implementing semantic-importance-based resource allocation in a simple manner, which illustrates a key advantage of TokenCom over conventional SemCom.

Beyond importance-based selection, efficiency can be further improved through token merging. Erak et al.~\cite{erakAdaptiveTokenMerging2025,erakAdaptiveParetoOptimalToken2025a} introduced adaptive token merging, a training-free framework that selectively merges semantically redundant tokens at the edge, effectively balancing inference speed and transmission costs. Devoto et al.~\cite{devotoAdaptiveSemanticToken2025,devotoAdaptiveSemanticToken2024} further designed an adaptive semantic token selection method based on DJSCC, which dynamically adjusts the number of tokens transmitted with end-to-end training, exhibiting excellent characteristics without the cliff effect.

Semantic importance can also be built into the token stream rather than estimated afterward. Coarse-to-fine tokenization~\cite{tianVisualAutoregressiveModeling2024} yields a reconstruction hierarchy. Among the 10 scales, the first 7 hold only 155 of the 680 indices but already support a coarse reconstruction of the entire input. This structure naturally supports layered transmission and unequal error protection (UEP): the receiver first decodes an early prefix and then improves the reconstruction when later scales are received. However, the protection order should still be validated based on distortion and downstream-task performance, since fine scales may carry critical details such as text or facial identity.

The second criterion, \textbf{contextual predictability}, is exploited by context-aware token masking~\cite{shinContextAwareWirelessToken2026}, in which the transmitter and the receiver share a masked language model (MLM). The transmitter greedily masks the position of lowest prior entropy under this MLM, updates the remaining priors after each step, and stops once masking one more token would reduce the overall detection probability. The receiver restores the masked positions with the same MLM. Since masked positions consume no channel resources, selection and power allocation collapse into one decision, as every masked token raises the energy available to the rest. The rule is also task-independent, being derived from a generic context model rather than from a task-specific attention map. Redundancy is thus removed against a model the receiver also holds, rather than at the encoder alone.

An autoregressive model exploits the same criterion in a coarser way. Zhang et al.~\cite{zhangAdaTokenComRateAdaptive2026} proposed Ada-TokenCom, which transmits only a prefix of the token sequence under arithmetic coding and lets the receiver generate the tail tokens with an identical pre-trained model. Since conditional self-information decreases along the generation order, later tokens are the more predictable ones, and the truncation length alone controls the rate, as the reconstructions in Figure~\ref{fig:tokenizer_compression} show.

Token-based selective transmission also extends naturally to additional modalities. The TaiChi framework~\cite{jiangTokenComVisionLanguageModel2026} introduced a dual visual tokenizer architecture: a ViT-based tokenizer captures global semantic structure at low resolution, while a convolutional tokenizer preserves high-resolution local details. A bilateral attention network (BAN) then fuses the two token streams, and a KAN-based projector aligns the resulting visual tokens with the text semantic space for joint VLM-channel coding. For video, Men et al.~\cite{menVideoTokenComTextual2026} proposed Video TokenCom, which uses a vision-language model to identify spatiotemporal tokens most relevant to the user's textual intent via optical-flow propagation. Relevant tokens receive full-precision encoding while the remainder are compressed, and a UEP-based adaptive source-channel coding strategy dynamically distributes bits according to semantic importance and channel conditions, achieving significant gains in perceptual quality at a lower source rate than both the diffusion-based video codec VC-DM~\cite{liExtremeVideoCompression2024} and H.265. For a representative MCL-JCV video quality assessment sequence~\cite{wangMCLJCVJNDBased2016} at 6~dB SNR, Video TokenCom maintains a PSNR of 28.82~dB whereas H.265 decreases to 11.24~dB, indicating that token-level recovery mitigates the cliff effect of conventional codecs. For scenarios where the data distribution shifts over time, Chen and Yang~\cite{chenEvolvingTokenCommunication2026a} proposed transmitting only a fixed-length prefix of each token's embedding, while the receiver employs a parametric memory network to reconstruct the full token, and this network is continuously updated online via a codebook-based distillation strategy. Compared with prior evolving-memory baselines, the method achieves up to 1.09~dB PSNR improvement under varying MIMO fading conditions.


\subsection{Token Transmission and Multiple Access at the Physical Layer} 

TokenCom can inherit and extend existing advantageous methods from SemCom at the physical layer. Ying et al.~\cite{yingJointSemanticChannelCoding2025} designed a joint semantic-channel coding and modulation (JSCCM) scheme specifically tailored for token streams. To handle varying semantic importance, Zhang et al.~\cite{zhangTokComUEPSemanticImportanceMatched2025} proposed TokenCom-UEP, which applies UEP based on the semantic significance of each token, outperforming equal protection schemes. These physical layer designs ensure reliable token transmission while preserving semantic integrity.

Going beyond static UEP, recent works exploit token semantics more deeply at the bit-mapping and resource-allocation levels. Lee et al.~\cite{leeSemanticsAwareHierarchicalToken2026} proposed a semantics-aware hierarchical token communication scheme (H-TokCom) that first groups tokens by semantic similarity using a language model, and then represents each token with a two-part codeword, specifically, a cluster prefix identifying its semantic category and a token suffix specifying its identity within the cluster. Higher transmit power is allocated to prefix bits, so that even when suffix bits are corrupted, the received token is likely to remain in the correct semantic cluster. This design achieves a 35.4\% improvement in semantic similarity (Sentence-BERT score~\cite{reimersSentenceBERTSentenceEmbeddings2019}, from 0.206 to 0.279) at 3~dB SNR on the COCO (Common Objects in Context) caption dataset. Liu and Wang~\cite{liuTONICTokenCentricSemantic2026} proposed token-centric semantic communication (TONIC), a framework that explicitly bridges the gap between bit-level link design and token-level task execution. At the transmitter, TONIC estimates the task relevance of each token and performs utility-aware UEP under a fixed channel budget. At the receiver, a confidence-gating module marks low-confidence decoded tokens as erasures, which are then recovered by a transformer-based completion model. TONIC consistently outperforms separation-based baselines, pixel-domain DeepJSCC, and other token-domain schemes on image classification tasks across additive white Gaussian noise (AWGN), Rayleigh, and Rician fading channels.

Semantic compression and link adaptation can also be optimized jointly. In Ada-TokenCom~\cite{zhangAdaTokenComRateAdaptive2026}, a Lyapunov-based cross-layer adapter selects the truncation length and the modulation and coding scheme (MCS) together under a long-term channel-symbol budget. The same coupling arises in the multi-user downlink studied by Zeinali et al.~\cite{zeinaliWirelessTokenComRLBased2026}, where the tokenizer choice fixes both the required rate and the achievable quality and is therefore optimized jointly with sub-channel assignment and beamforming. The scheme reduced video freezing events by about 68\% against an H.265 baseline for 1080p video with 16 users. The two works adapt the rate at different granularities, one switching the tokenizer and the other truncating within a fixed one.

\begin{table*}[!t]
\centering
\caption{Sources of the token priority weight $w_i$, and the decision each has been used to drive. Some weights are defined on feature channels or on tokenizer scales rather than on individual tokens}
\label{tab:token_weight} 
\scriptsize
\setlength{\tabcolsep}{2pt}
\begin{tabular}{|>{\raggedright\arraybackslash}p{5.9cm}|>{\raggedright\arraybackslash}p{2.3cm}|>{\raggedright\arraybackslash}p{3.2cm}|>{\raggedright\arraybackslash}p{5.0cm}|}
\hline
\textbf{How $w_i$ is obtained} & \textbf{Representative work} & \textbf{Decision driven in that work} & \textbf{Limitation} \\
\hline
Attention a token receives from the other modalities, averaged and ranked to retain the top $k\%$ & LMECM-SC~\cite{pengLargeModelEmpowered2025} & Token selection & Measures cross-modal salience rather than task value. Demonstrated as a way of reducing training tokens \\
\hline
Contextual predictability, read inversely from the prior entropy of each position under an MLM shared by transmitter and receiver, so that low-entropy tokens are masked & Shin et al.~\cite{shinContextAwareWirelessToken2026} & Whether to transmit or mask & Task-independent by construction, and therefore carrying no information about task value \\
\hline
Semantic loss caused by losing each frame, or the number of important words it contains, scored by a pre-trained language model & SIAC~\cite{guoSemanticImportanceAwareCommunications2023} & Power allocation & Defined for text only \\
\hline
Downstream task contribution, taken from the gradient of the correct-class probability with respect to each feature channel & IRCSC~\cite{sunTaskOrientedSemanticCommunication2025} & How many semantic features to retain & Defined per feature channel rather than per token, and only locally valid as a linear approximation \\
\hline
Tokenizer structure, in which coarse scales are decoded first and finer scales encode residuals conditioned on them & VAR~\cite{tianVisualAutoregressiveModeling2024} & Coarse-to-fine generation order & Free of cost, but that work performs no transmission, and the scale order need not follow task importance \\
\hline
Feature-wise masking, applied to a representation whose importance order has been reshaped by decreasing retention probabilities and increasing distortion levels across the feature index & ISFR~\cite{zhanTowardRobustSemantic2026} & Channel matching, feature selection, modulation and power & Requires retraining the semantic encoder \\
\hline
\end{tabular}
\end{table*}

The designs above address different parts of a single resource allocation problem. Let $w_i \ge 0$ denote the transmission priority of the $i$-th token, obtained from either criterion of Section~\ref{subsec:compress} and listed in Table~\ref{tab:token_weight}, and let $D_i$ be the semantic distortion incurred when that token is sent with power $p_i$, modulation order $m_i$ and coding rate $r_i$. A token-aware physical layer then solves
\begin{equation} 
\min_{\{p_i, m_i, r_i\}} \sum_i w_i D_i(p_i, m_i, r_i), \quad \text{s.t.} \sum_i p_i \le P, \quad \sum_i \frac{s_i}{r_i \log_2 m_i} \le C,
\end{equation}
where $s_i$ is the bit length of token $i$, $P$ the transmit power budget and $C$ the symbol budget. Zhan et al.~\cite{zhanTowardRobustSemantic2026} minimized this importance-weighted distortion jointly over channel matching, feature selection, modulation and power, and Park et al.~\cite{parkImportanceAwareSemanticCommunication2026} solved its quantization, subcarrier mapping and power version for MIMO orthogonal frequency division multiplexing (OFDM) systems, taking $w_i$ from the attention scores of a pre-trained ViT. The paragraphs below follow $w_i$ into the power, modulation and rate variables of this problem, and then into the retransmission decision that sits on top of it, and then to the multiple-access case in which the channel itself is shared.

The mapping from importance to power predates TokenCom. Guo et al.~\cite{guoSemanticImportanceAwareCommunications2023} scored the semantic importance of each data frame with a pre-trained language model embedded in a cross-layer manager, and allocated power to minimize the importance-weighted outage probability. Xu et al.~\cite{xuDataImportanceAwarePowerAllocation2025} derived a closed-form waterfilling solution for the analogous importance-weighted problem, in which the power assigned to a sub-stream grows with the logarithm of its importance weight and is scaled by the inverse of its channel gain. Above 10~dB SNR it lowers the normalized importance-weighted mean squared error by over 7~dB and 10~dB against margin-adaptive waterfilling and equal power allocation under their joint sub-pixel and segment-level data partitioning, by 4~dB and 4.5~dB under sub-pixel-level partitioning alone, and by 2.5~dB and 5.2~dB under segment-level partitioning alone, the gain being largest where the importance values vary most widely. The importance weight enters this solution through a logarithm, so the optimum is a power offset rather than a power ratio: a token ten times as important does not warrant ten times the power. The offset is also inversely proportional to the channel gain and the allocation is truncated at zero, so on a sufficiently poor link the low-importance tokens receive no power at all. In TokenCom such tokens can be left to receiver-side generation, at which point the power allocation problem reduces to the token selection problem of Section~\ref{subsec:compress}.

Modulation carries the same weight into the geometry of the constellation. Conventional design makes every symbol error equally unlikely, whereas semantic-aware design makes the errors that do occur as harmless as possible. Shaju et al.~\cite{shajuNotAllSymbols2026} learned a QAM constellation whose symbol positions follow the co-occurrence statistics and task-relevance scores of VQ-VAE concepts, and proved that any Gray-coded $M$-QAM constellation is strictly suboptimal under their criticality-weighted symbol vulnerability criterion, which weights each symbol by both its task-relevance score and its co-occurrence probabilities, whenever the source exhibits non-uniform semantic importance and co-occurrence statistics.

Rate selection couples token selection to link adaptation, as the joint truncation and MCS adaptation of Ada-TokenCom above illustrates: a good channel carries more tokens under a more bandwidth-efficient scheme, while a poor one truncates the sequence and leaves more tail tokens to the receiver's autoregressive model. There the candidate set is formed by pairing five coding options, uncoded transmission among them, with quadrature phase shift keying (QPSK) and 16-QAM, and each option is characterized by a pair of offline-fitted coefficients giving a waterfall SNR threshold that grows logarithmically with packet length, from which the packet error rate follows in closed form. That adaptation is driven by the channel rather than by $w_i$. Sun et al.~\cite{sunTaskOrientedSemanticCommunication2025} instead let semantic importance and channel condition jointly set how many semantic features to send, and proposed a semantic transmission integrity index that relates the amount of correctly delivered semantics to inference performance.

Retransmission is where the receiver-side generative prior changes the set of available actions. Hu et al.~\cite{huSemHARQSemanticAwareHybrid2026} replaced the CRC trigger of hybrid automatic repeat request (HARQ) with a feature distortion evaluation network that assesses the corruption of each semantic feature, retransmitting only those judged corrupted and using the remaining channel resources to send further features in descending order of importance, raising rank-1 accuracy in vehicle re-identification by more than 20\% and vehicle color classification accuracy by 10\% over earlier schemes in the low SNR regime. Han et al.~\cite{hanGenerativeSemanticHARQ2026} ran a transformer-based variational autoencoder as a lightweight overlay on the conventional stack, whose stochastic encoder draws a different latent representation at every attempt, so that a single model supplies incremental knowledge without a protocol designed to provide it. Since the receiver holds a generative prior, regeneration is itself an available action, so the decision becomes a choice among retransmitting, regenerating from context and discarding.

Receiver-side processing also reshapes uplink multiple access. In the token-domain multiple access (ToDMA) scheme of Qiao et al.~\cite{qiaoToDMALargeModelDriven2025}, devices share a tokenizer and a modulation codebook whose columns are indexed by the token, so a transmitted signature identifies the token rather than the device and access needs no grant. The receiver detects the active token indices in each slot by compressed sensing and estimates their channels, then clusters those channels across the block into device streams, which is possible because a device's channel stays approximately constant over the sequence. The clustering separates streams without recovering device identity, and devices that send the same index in one slot collapse into a single detection, leaving positions in the recovered sequences unfilled. The physical layer hands each such position a candidate set rather than a decision: a single candidate is filled directly, and only an ambiguous one is resolved by the masked-token prediction discussed in Section~\ref{subsec:robust}. At 20 active devices the TER stayed below that of the same scheme with contextual recovery removed across 0 to 25~dB, and at 10~dB the margin grew as the number of devices increased to 80.

\subsection{Robust Token Transmission and Error Correction}\label{subsec:robust} 

Beyond reducing transmission overhead, tokenization provides a new reliability mechanism: missing or unreliable tokens can be inferred from their surrounding context, cross-modal side information, and channel observations. Such receiver-side semantic recovery can reduce the reliance on packet retransmission, which is particularly valuable in real-time or long-delay links where retransmission is costly or impractical~\cite{huTokenEncodingSemantic2026,wangResiCompLossResilientImage2025}. Unlike conventional error correction that targets exact bit recovery, however, LM-based recovery is more accurately regarded as semantic-level error concealment or recovery, with retransmission retained for low-confidence cases.

Early works on discrete codebook-based SemCom laid the groundwork for token-level error resilience. For example, Hu et al.~\cite{huRobustSemanticCommunications2023} proposed a robust SemCom system based on masked VQ-VAE. By quantizing semantic features with a discrete codebook and only transmitting codebook indices, it significantly reduces communication overhead. They also introduced adversarial training and weight perturbation mechanisms to enhance the system's robustness to semantic noise. This method substantially improved interference resistance in tasks, but did not fully exploit cross-modal semantic information and generative model contextual reasoning. When tokens are discrete, systems can directly leverage the generative capabilities of pre-trained LMs for semantic-level error concealment, rather than relying on a dedicated corrector as in earlier text SemCom systems~\cite{pengRobustSemanticText2024}. Liu et al.~\cite{11149073} proposed a text-guided TokenCom system that represents images as token sequences and uses text descriptions as side semantic information. At the receiver, a masked token generation model restores lost tokens, validating the token's effectiveness in semantic error correction. Advanced receiver designs are also leveraging the discrete nature of tokens. Shin et al.~\cite{shinContextAwareWirelessToken2026} proposed a joint token masking and detection framework in which transmitter and receiver share an MLM as a common context prior. The receiver performs Bayesian detection that combines the channel likelihood with the MLM's token prediction distribution, while the transmitter masks tokens that the receiver can confidently infer from context, achieving up to 1.77$\times$ and 1.63$\times$ reconstruction performance gains on the Europarl and WikiText-103 datasets, respectively. Hu and Li~\cite{huTokenEncodingSemantic2026} proposed TokCode, a token re-encoding scheme that transforms a token sequence into a same-length sequence drawn from the same codebook, but with an arrangement that is significantly more resilient to token loss. To avoid expensive end-to-end retraining, they introduce soft foundational model adaptation (SFMA) to align a pre-trained generative foundation model with the re-encoded token space. In prompt-driven generative image transmission experiments, TokCode preserves semantic fidelity even when 40--60\% of tokens are randomly erased during transmission.

Recent work further explores token-level resilience. Wang et al.~\cite{wangResiCompLossResilientImage2025} proposed a dual-functional masked visual token modeling framework that unifies entropy modeling and feature-domain packet loss concealment, significantly improving the resilience of neural image codecs under packet losses.

\subsection{Token Communication for LM Services}\label{subsec:lmsvc} 

While Section~\ref{subsec:Collaboration} presents edge-device collaboration as a deployment strategy to support LM-driven SemCom, this section focuses on another perspective: token transmission as an intrinsic infrastructure requirement of LM inference itself, regardless of the specific semantic task.

\begin{figure*}[!tb]
    \centering
    \subfloat[]{%
        \includegraphics[width=\textwidth]{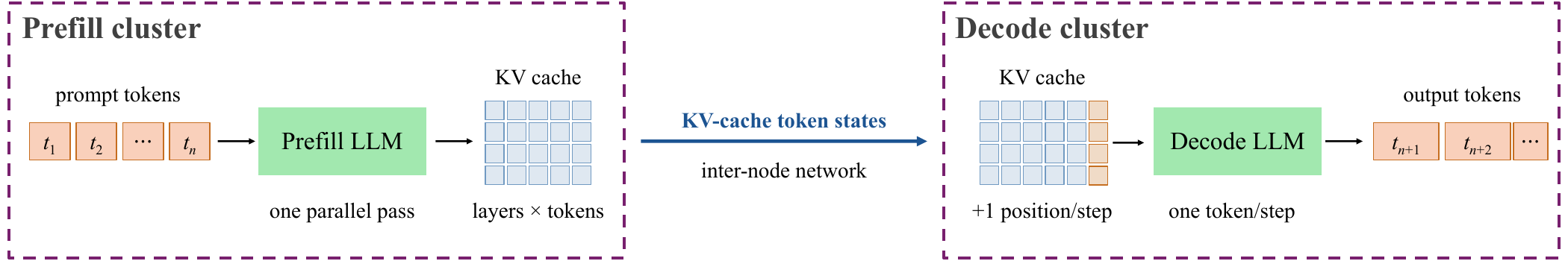}%
        \label{fig:lmsvc_a}}\par\smallskip
    \subfloat[]{%
        \includegraphics[width=\textwidth]{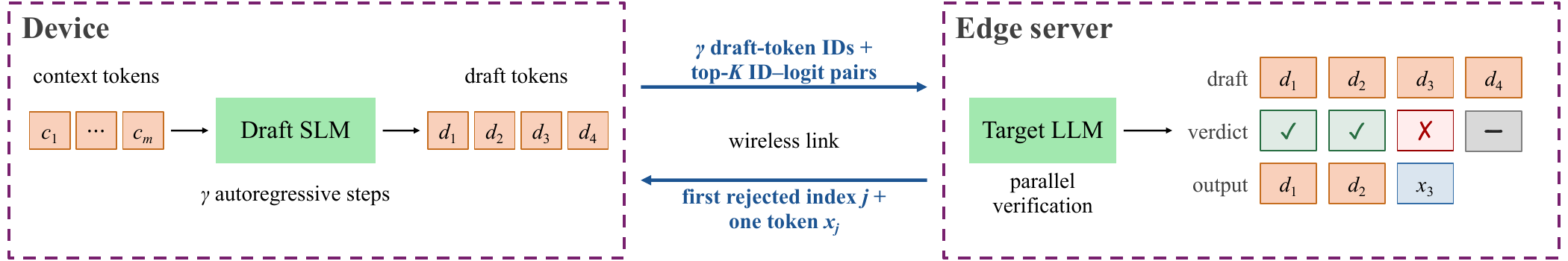}%
        \label{fig:lmsvc_b}}\par\smallskip
    \subfloat[]{%
        \includegraphics[width=\textwidth]{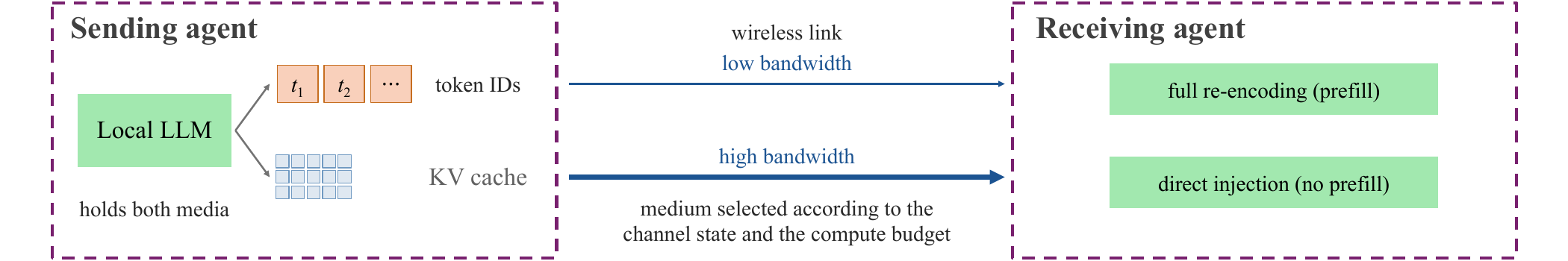}%
        \label{fig:lmsvc_c}}
    \caption{Token communication inside LM service infrastructure.
(a) KV-cache transfer in disaggregated LLM serving. The prefill cluster processes the prompt and transfers the resulting KV cache to the decode cluster, which generates tokens autoregressively while extending the cache at each step. The link depicts a cross-node deployment~\cite{mooncake2024,distserve2024}.
(b) Token-level speculative decoding over a device--edge wireless link, drawn
for $\gamma = 4$ with the third draft rejected. The uplink carries the $\gamma$
draft identifiers and their per-step top-$K$ logits. The target LLM at the edge
verifies all drafts in one parallel pass, keeps the accepted prefix up to the
first rejection at index $j$, discards the drafts after it, and returns only $j$
and one token $x_j$ ($j = \gamma+1$ when all are
accepted)~\cite{zhengCommunicationEfficientCollaborativeLLM2025}.
(c) Communication-media selection over a wireless link. Because tokens are
generated during inference, the sending agent holds both media at transmission
time, and neither dominates: the token identifiers cost little bandwidth but
oblige the receiver to re-encode the context with a full prefill pass, whereas
the KV cache skips the prefill at a much higher bandwidth
cost~\cite{daiTokenKVCacheCommunication2026}.}
    \label{fig:lmsvc} 
\end{figure*}

Beyond semantic source coding, tokens are also the basic units of communication in LM service infrastructure. In MoE architectures, token representations are routed to experts distributed across devices, typically through collective communication such as all-to-all, which can become a major scalability bottleneck~\cite{megascaleMoE2025}. In disaggregated LLM serving systems, the prefill and decode stages are deployed on separate GPU instances or clusters, requiring per-request KV cache token states to be transferred between them. Consequently, KV-cache placement and network bandwidth become key design considerations for meeting end-to-end latency requirements~\cite{mooncake2024,distserve2024}. Figure~\ref{fig:lmsvc_a} illustrates this disaggregated serving pipeline. These examples show that TokenCom is not merely a concept at the semantic layer but a foundational infrastructure requirement for LM-native AI services, with the network becoming an integral part of the inference pipeline itself. Although KV-cache tensors are not tokens themselves, their dimensions and transmission patterns are directly determined by the token sequence and therefore constitute token-associated communication payloads in distributed LM serving.

A key advantage of using tokens as the fundamental transmission unit is that it makes speculative decoding, a technique originally developed for accelerating autoregressive LM inference in computer systems~\cite{leviathanFastInferenceTransformers2023}, viable in wireless settings. The key observation is that token-level granularity provides the right level of abstraction for device-edge collaboration: it is neither too coarse (as with feature vectors) nor too fine (as with individual bits), but lies at the semantic-unit level where meaningful draft-and-verify collaboration can occur.

With token as the transmission unit, a draft SLM at the device can rapidly generate $\gamma$ draft tokens, which are then verified in parallel by a target LLM at the edge server~\cite{zhengCommunicationEfficientCollaborativeLLM2025}. Because the edge retains the context, each round only needs to carry $\gamma$ draft identifiers and their per-step top-$K$ logits on the uplink, while the downlink returns a single token. The top-$K$ truncation is applied at the SLM's own sampling step, so the transmitted sparse set is the draft distribution itself and the verified output distribution remains identical to that of the target LLM, at the cost of a modified acceptance rate. Transmitting these truncated logits instead of the entire vocabulary distribution is what keeps the uplink cost low while maintaining semantic fidelity. Only token-level semantics allow meaningful draft generation and efficient verification. Figure~\ref{fig:lmsvc_b} depicts the corresponding device-edge draft-and-verify loop.

Beyond speculative decoding, tokens also enable distributed inference across multiple devices. In IoE scenarios, terminal devices with limited computational capabilities can leverage tensor parallelism to offload token inference across multiple edge devices, using over-the-air aggregation for efficient coordination~\cite{zhang2505ZuHuiCommunicationEfficientDistributedOnDevice2025}. For complex multimodal scenarios, Zhang et al.~\cite{zhangTaskOrientedMultimodalToken2026} further investigated task-oriented multimodal token transmission, focusing on cross-modal alignment to optimize fusion efficiency in resource-constrained networks.

Tokens are also becoming the native communication medium for multi-agent LLM systems, and represent an emerging direction for TokenCom beyond point-to-point transmission. Chen et al.~\cite{chenScalingVideoUnderstanding2026} proposed a compact latent multi-agent collaboration framework (MACF) for long video understanding, where each local agent encodes its observations into compact tokens within a shared embedding space, and a central coordinator performs global reasoning over these agent tokens. Curriculum learning progressively strengthens semantic alignment, evidence summarization, and cross-agent coordination. At the protocol level, Mou et al.~\cite{mouHyLaTEfficientMultiAgent2026} proposed HyLaT, a hybrid latent-text protocol for LLM multi-agent systems that routes elaborate cognitive signals through a latent channel for communication efficiency while preserving natural language for concise critical signals such as answers, commitments, and final decisions. A two-stage training procedure, comprising single-agent hybrid generation learning followed by multi-agent interactive co-training, enables the protocol to generalize across diverse cooperative tasks. At the infrastructure level, Dai et al.~\cite{daiTokenKVCacheCommunication2026} addressed the joint optimization of communication media selection and resource allocation in multi-agent collaboration: since transmitting discrete token identifiers and transmitting KV cache states incur different computation and communication costs, the proposed low-complexity joint media selection and resource allocation (JMSRA) algorithm adaptively selects the communication medium and allocates bandwidth according to the channel conditions and per-agent computation budgets, reducing end-to-end collaboration latency. Figure~\ref{fig:lmsvc_c} contrasts the two media and the receiver-side cost each of them entails.

Which of the two media is preferable depends on the link. Lee et al.~\cite{leeLowLatencyEdgeLLM2026} compared them for edge LLM handover: forwarding the past tokens and re-running prefill incurred a delay independent of the backhaul rate, whereas the delay of transferring the KV cache fell as that rate rose. Their cache-based scheme was therefore more than 3.1 times slower than their joint design at a backhaul rate of 2~Gbps, and converged with it at high rates. The payload gap drives this behavior: Chen et al.~\cite{chenFederatedInferenceHeterogeneous2026} reported 88~KB per token for cache exchange against 16 bytes for text-based exchange, both aggregated over four sharer models, in a setting built on Cache-to-Cache~\cite{fuCachetoCacheDirectSemantic2025}, at an accuracy gain of about 15\%. Reliability separates them further: a token stream can be repaired from the receiver's generative prior, as Section~\ref{subsec:robust} describes, so the two media may be operated at different residual error rates at the same SNR. Token identifiers are therefore the better medium at low SNR. The prefill cost borne by the token path is set by the receiver compute and the context length, and does not grow as the channel degrades, whereas the transmission cost borne by the cache path grows without bound. The token stream can also trade residual errors for rate where the cache cannot. Table~\ref{tab:media_comparison} contrasts the two media, the cache side is developed further in~\cite{yeKVCOMMOnlineCrosscontext2025,yangSpatialPrefixCaching2026}, and latent communication more broadly is surveyed in~\cite{liuBeyondTokensUnified2026}. Cache transfer remains the right choice within a cluster and over high-rate backhaul, where the ordering reverses. These works collectively establish token routing and token protocol design as core research problems in the LM-native communication layer.

\begin{table*}[!t]
\centering
\caption{Token identifier transmission versus KV-cache transmission as the communication medium between two LM endpoints. Payload values are those reported by Chen et al.~\cite{chenFederatedInferenceHeterogeneous2026} for a four-sharer cache collaboration setting. $\lvert \mathcal{V} \rvert$: vocabulary size; $N_{\mathrm{layer}}$: number of layers; $N_{\mathrm{kv}}$: number of key-value heads; $d_{\mathrm{head}}$: head dimension; $b$: bits per element}
\label{tab:media_comparison} 
\small
\setlength{\tabcolsep}{4pt}
\begin{tabular}{|>{\raggedright\arraybackslash}p{3.0cm}|>{\raggedright\arraybackslash}p{6.4cm}|>{\raggedright\arraybackslash}p{6.4cm}|}
\hline
\textbf{Aspect} & \textbf{Token identifier transmission} & \textbf{KV-cache transmission} \\
\hline
Per-token payload & $\log_2 \lvert \mathcal{V} \rvert$ bits for the identifier; 16 bytes for the text-based exchange measured in~\cite{chenFederatedInferenceHeterogeneous2026} & $2 N_{\mathrm{layer}} N_{\mathrm{kv}} d_{\mathrm{head}} b$ bits per model; 88~KB for four sharers combined in the same experiment \\
\hline
Receiver-side cost & One prefill pass over the received context & No prefill for same-model transfer. Under Cache-to-Cache the receiver still prefills its own input and adds a projection and fusion pass~\cite{fuCachetoCacheDirectSemantic2025} \\
\hline
Cross-model use & A shared tokenizer or tokenizer negotiation as in~\cite{zeinaliWirelessTokenComRLBased2026} & Token-level and layer-level alignment plus a fuser trained for the model pair~\cite{fuCachetoCacheDirectSemantic2025} \\
\hline
Effect of residual errors & Index errors can be concealed using the receiver's generative prior over token sequences (Section~\ref{subsec:robust}) & Perturbed tensor entries enter the attention computation. Task-level concealment remains an open problem \\
\hline
Retransmission & Held at one transmission per packet in~\cite{qiaoTokenCommunicationsLarge2025} by concealing erasures at the receiver & Relies on automatic repeat request (ARQ) or a lower code rate to preserve tensor fidelity \\
\hline
Privacy & The exchanged text is directly readable & Not directly readable, though the cache still encodes the source content \\
\hline
Favorable regime & Low link rate; low SNR; receiver computation available & High link rate; receiver computation constrained, or time to first token critical \\
\hline
\end{tabular}
\end{table*}

\subsection{Token Communication for Embodied and Agentic Intelligence} 

In the scenarios examined so far, whatever feedback the receiver produces stays inside the information exchange itself: a decoder, an inference engine, or a peer agent replies with further tokens, and nothing it does alters the physical world from which the transmitter draws its next input. Embodied AI, a setting also studied under the name physical AI~\cite{wangWorldModelEnabledCausal2026}, removes that separation, since in agents such as robots, autonomous vehicles, and extended reality devices the decoded tokens drive actuators, the actuators change the environment, and the environment supplies the next observation~\cite{sajidEmpoweringEmbodiedAI2026}. Closing the loop through the physical world extends TokenCom in three connected ways: it changes what a token error costs, what a token can represent, and what makes a token worth sending.

The first change concerns what a token error costs. Sajid et al.\ argued that a network carrying such traffic is part of the control loop rather than merely a transport layer beneath it, since a link that meets its packet error and throughput targets can still leave a control loop unstable, or leave an automated vehicle acting on perception that arrived too late. A transmission error is then no longer an isolated distortion: an incorrect decision changes the state that produces the next observation, so the error accumulates along the trajectory. This is a second propagation path alongside the one inside autoregressive generation, which travels along the token dependencies within a single generated sequence. The two act in series when a corrupted token distorts the generated action sequence and the resulting action then perturbs the environment. Receiver-side regeneration still helps while an action is pending, but not once it has been executed. Two studies indicate the resulting design target. Dai et al.~\cite{dai6GCommunicationNetworks2026} connected a haptic device, an industrial robotic arm, and an intermediary platform over a 5G open radio access network testbed, and measured millisecond latency under stable closed-loop operation. Wang et al.~\cite{wangSemanticbasedInternetEmbodied2026} reported a simulated robotic grasping study in which a pipeline based on bit-level image reconstruction failed when the SNR dropped, whereas semantic transmission maintained task success with substantially lower end-to-end latency, so that the comparison is settled by task success rather than by a link-level metric.

The second change concerns what a token can represent. The shared semantic space of Unified-IO 2~\cite{luUnifiedIO2Scaling2024}, introduced at the beginning of this section as evidence for multimodal unification, already covers robot actions and robot state alongside text, images, and audio. In an embodied system that coverage becomes a property of the link rather than of the model alone. Among the vision-language-action (VLA) models surveyed by Ma et al.~\cite{maSurveyVisionLanguageAction2026}, those built on a discrete action head make the construction concrete. Following the scheme such models introduced, Kim et al.~\cite{kimOpenVLAOpenSource2025} discretized each dimension of a continuous robot action into 256 bins and mapped them to the 256 least frequently used entries in the underlying language model vocabulary, so that their 7B-parameter model, trained on 970k real-world robot demonstrations, generates actions with the same next-token objective used to generate text. An action is thus representable as a short sequence of tokens drawn from the same vocabulary as text, which is enough to change the communication architecture: one token interface can carry both the perception uplink and the action downlink, whereas a conventional design pairs a video codec on the uplink with a distinct control message format on the downlink. It also binds the two halves of the loop, since Unified-IO 2 encodes the robot state with the same token inventory that carries the action commands, so state reports and commands can be scheduled against one budget.

The third change concerns what makes a token worth sending, and here the argument widens from a single embodied agent to a group of cooperating agents, not all of which need have a physical body~\cite{zhangTowardsSemanticbasedAgent2026}. The criteria of Section~\ref{subsec:compress} judge a token by what it contributes to the current transmission, whether that is reconstruction quality, the immediate task, or what the receiver could have inferred without it. A closed loop leaves contextual predictability intact, but it changes the horizon over which semantic importance is evaluated and adds a further consideration that neither criterion covers. Wang et al.~\cite{wangWorldModelEnabledCausal2026} made the change of horizon explicit, formulating semantic transmission for closed-loop systems as the maximization of long-term return per bit and scoring each token by the difference in expected long-horizon return between transmitting it and not transmitting it, an estimate obtained from imagined rollouts of a world-model-based causal digital twin. The added consideration concerns the receiver's internal state rather than the content of the message. Seo et al.~\cite{seoReasoningNativeAgenticCommunication2026} observed that two agents can recover the meaning of a message identically and still act in conflict, because their internal reasoning processes have evolved differently. They termed the resulting mismatch belief divergence and placed a reasoning coordination plane above the data delivery plane, which predicts how far the two internal states have drifted and decides on that basis whether a transmission is needed at all. The two mechanisms answer different questions, which tokens a message should carry in the first case and whether to transmit at all in the second, but both replace a per-transmission criterion with one defined by the task that the agents are carrying out.

\subsection{Token-Level Protocol Design}\label{subsec:protocol} 

Above the physical and access layers, a token stream must still be packetized, framed, acknowledged, ordered, and rate-controlled, but a generative receiver changes how each of these functions is performed. Deployment-oriented designs accommodate these changes within the layered stack. Zhang et al.~\cite{zhangTowardsNativeAI2026} mapped semantic processing onto the existing mobile protocol stack while preserving its structure. They added semantic extraction and coding at the application layer, semantic importance identifiers at the data link layer to support semantic-aware CRC, HARQ, and resource block allocation, and importance-driven coding and modulation scheduling at the physical layer. Tung et al.~\cite{tungMultilevelReliabilityInterface2025} reached a similar conclusion based on a deployment constraint: application and network providers are separate entities connected over general-purpose transmission control protocol/Internet protocol (TCP/IP) links, leaving a fully joint design without timely channel state at the source. They therefore designed the source and channel mappings separately and connected them through a multi-level reliability interface.

As Lee et al.~\cite{leeSemanticPacketAggregation2025} observed, packetization reduces to a choice of packet length when all bits are treated equally, and becomes a grouping problem once tokens carry unequal and mutually dependent semantics. They formalized that problem as maximizing the expected similarity between the token subset that survives an erasure channel and the original message, and showed that the groups must be optimized jointly because the value of one packet depends on which others arrive, and they obtained near-optimal groupings at a fraction of the cost of exhaustive search~\cite{leeLowComplexitySemanticPacket2025}. Qiao et al.~\cite{qiaoTokenCommunicationsLarge2025} addressed the same interdependence from the transmitter side, randomizing the token positions carried in each packet so that a loss removes scattered rather than consecutive tokens. Their reason is that burst errors significantly degrade reconstruction, because a run of consecutive losses takes with it the context needed to infer the missing tokens.

Framing must carry a field whose content is set by the shared model rather than by the channel code. When the transmitter omits tokens that the receiver can infer, as in the context-aware masking of Section~\ref{subsec:compress}, the receiver's detection rule depends on which positions were masked, and the per-token power that the transmitter applies and the receiver's likelihood assumes depends on how many~\cite{shinContextAwareWirelessToken2026}. The masked set must therefore be signaled, and its overhead must be included in the rate budget. Control signaling changes for the same reason. For example, the downlink control information format of Wang et al.~\cite{wangHybridSemanticRAN2025} jointly indicates semantic and conventional transmissions, allowing both to share one physical downlink shared channel.

Where Section~\ref{subsec:robust} covers the recovery algorithms themselves, the protocol question is what triggers them and at what granularity. Retransmission changes in three ways: the trigger shifts from a failed CRC to insufficient semantic confidence, the granularity shifts from a transport block to individual tokens or token groups, and combining moves from the bit domain to the latent domain. Hu et al.~\cite{huSemHARQSemanticAwareHybrid2026} realized the first two changes at the feature level through per-feature distortion evaluation, and Han et al.~\cite{hanGenerativeSemanticHARQ2026} the third through a quality-aware combiner that merges the received latent vectors. Both schemes define a feedback signal of their own, a per-feature binary retransmission request in the former and a quality-triggered one in the latter, but neither is tied to the sequence numbering and window management on which selective repeat is built. TONIC~\cite{liuTONICTokenCentricSemantic2026} provides the corresponding receiver-side mechanism through confidence gating followed by erasure completion.

Choosing among retransmission, regeneration, and discarding requires trading off quantities measured in different units: channel occupancy, inference latency, and residual semantic distortion. Qiao et al.~\cite{qiaoTokenCommunicationsLarge2025} quantified both sides of this trade-off for image transmission. Predicting erroneous positions from cross-modal context instead of requesting them again reduces the average number of transmissions per packet from $T = 1/(1-\mathrm{PER})$ under a repeat protocol to one. However, receiver-side tokenization and context processing are estimated at 19.2 and 34.9~ms per image from the nominal throughput of two contemporary edge accelerators, whereas the communication time per image, which already accounts for the average number of retransmissions at each operating point, falls from 43.7 to 26.7~ms as the SNR rises from 6 to 9~dB. For this workload, computation replaces retransmission only once the channel is poor enough, and the surveyed studies offer no general online policy for making that choice.

Ordering constraints originate in the tokenizer and must then be realized over the transport. Autoregressive decoding conditions each token on its predecessors, so reordering changes the distributions from which the output is drawn rather than simply permuting it. However, strict sequencing is more than some tokenizers require. In a coarse-to-fine construction~\cite{tianVisualAutoregressiveModeling2024}, a scale can be decoded only after every coarser scale has arrived, whereas the indices within one scale are predicted in a single pass and may arrive in any order. Such a tokenizer defines a partial order that lies between the total order enforced by TCP and the absence of ordering in the user datagram protocol (UDP). Neither exposes that partial order natively.

Context synchronization is essential to all of the above because the reliability of a generative receiver depends on how closely its context remains aligned with the transmitter's. Mobility disrupts this alignment. Lee et al.~\cite{leeLowLatencyEdgeLLM2026} showed that restoring it after a handover is itself a scheduling problem: the target edge server can rebuild the context by forwarding tokens and re-running prefill or fetch it as KV cache over the backhaul, and the split between the two determines the delay. Between agents the same requirement extends from the token stream to the internal state behind it. Seo et al.~\cite{seoReasoningNativeAgenticCommunication2026} observed that two agents may interpret a message identically yet act inconsistently once their internal reasoning states drift apart, a condition they named belief divergence. They proposed triggering transmission based on the predicted misalignment of those states rather than on channel conditions or data novelty alone. Their architecture places a reasoning coordination plane above a conventional data delivery plane, and its policy of withholding a token that the counterpart can already infer is analogous to context-aware masking on a single link.

Congestion control is the least developed of the functions surveyed here. A conventional controller regulates one quantity, the offered load, which it reduces when loss or delay signals congestion. A TokenCom sender has a second option: it can lower the rate while increasing the share of the message the receiver generates, shifting completion of the message from the channel to the receiver's generative model. Qiao et al.~\cite{qiaoTokenCommunicationsLarge2025} outlined the qualitative implications: a context-bearing stream tolerates a less reliable transport such as UDP and permits relaxed flow and congestion control. Quantitative treatments remain limited to a single link, adapting the source compression rate and the modulation and coding scheme~\cite{zhangAdaTokenComRateAdaptive2026} or the drafting budget of the speculative decoding of Section~\ref{subsec:lmsvc}~\cite{tangGELATOGenerativeEntropy2026} through Lyapunov optimization. Extending either approach to several token streams sharing a bottleneck raises a fairness question that a bit-level controller cannot pose, because how much rate a flow can give up depends on the generative prior at its receiver, a quantity the network does not observe. Whether protocols for this regime should be designed or, following the protocol-learning and language-oriented approaches surveyed by Park et al.~\cite{parkTowardsSemanticCommunication2024}, learned from the task itself remains open.

\section{Challenges and Future Research Directions}\label{sec:Challenges} 

The evolution from SemCom to TokenCom marks a paradigm shift from fragmented semantic representation to unified token-native semantic transmission. By integrating LM-driven multimodal token processing, TokenCom enables high semantic compression, cross-modal context awareness, robust transmission, and low-latency communication, offering a foundation for a unified intelligent communication framework. Despite its theoretical and experimental success, LM-based intelligent communication still faces several critical challenges. These challenges can be categorized into two types: 1)~\textbf{general challenges} inherited from LM-driven SemCom that remain critical for TokenCom deployment, and 2)~\textbf{TokenCom-specific challenges} that arise uniquely from the token-based paradigm.

\subsection{General Challenges for LM-Driven Communication} 

\subsubsection{Privacy and Security Risks} 

Privacy and security remain primary concerns yet are rarely addressed in current research.
Future networks will process user semantics in addition to raw bitstreams.
Tokens, which serve as compact semantic carriers, may therefore inadvertently expose user intentions, behavioral patterns, or sensitive content.
Attackers could infer private information by intercepting a small token sequence or manipulate semantic meaning by altering a few tokens. Conventional encryption protects a serialized token stream in transit. The exposure arises instead at endpoints that legitimately hold the plaintext tokens, where semantic-level privacy and access control have no counterpart in current schemes. Future research should explore differential privacy mechanisms tailored to user semantics, encryption within token embedding space, and lightweight semantic-aware encryption algorithms that preserve both privacy and reconstruction efficiency.

\subsubsection{Model Complexity and Inference Latency} 

Model size and inference delay remain bottlenecks for real-time TokenCom deployment. Although TokenCom reduces retransmission overhead through contextual awareness, its reliance on multimodal LMs incurs substantial computational latency, challenging low-latency requirements at the edge. A promising direction is to establish collaborative inference frameworks that deploy lightweight models on devices while dynamically offloading complex reasoning tasks to edge or cloud servers. Techniques such as speculative decoding can further balance inference efficiency and communication delay. Efficient model compression and lightweight architecture design will also be essential to improve processing speed without sacrificing performance, and to keep the resulting energy cost within the sustainability targets set for 6G~\cite{youWhenAIMeets2025}.

\subsubsection{Cross-Model Compatibility} 

An LM-driven link couples two models, and the receiver can interpret a semantic representation only insofar as its own model assigns it the same meaning. That meaning is induced by architecture, training corpus and alignment procedure, so endpoints from different vendors cannot be assumed to interoperate, nor can a deployed terminal be assumed to remain compatible with a network-side model that has since been retrained. A 6G network must therefore negotiate model capability at connection setup, as it already negotiates radio access capability, and keep the agreed interface valid across model updates, either by constraining what an update may change or by supplying a mapping between the representation spaces on either side of it.

\subsubsection{Robustness and Adversarial Resilience} 

Pretrained LMs may inherit biases from training data or produce misleading semantics under adversarial perturbations. Enhancing TokenCom robustness requires integrating debiasing strategies, adversarial training, and interpretability analysis to strengthen semantic reliability and security in open environments. Such remedies presuppose that the failures can be detected. Benchmarks for large multimodal models already measure hallucination, fairness and bias, and robustness to adversarial prompts and perturbed inputs~\cite{zhangLargeMultimodalModels2025}, but a TokenCom receiver would have to raise the same alarms on the reconstruction it is generating rather than on an offline benchmark suite.

\subsection{TokenCom-Specific Technical Challenges} 

While the above challenges apply broadly to LM-driven SemCom, TokenCom introduces additional unique technical issues that stem directly from its token-based paradigm. These challenges require dedicated research attention and novel solutions tailored to the semantic nature of tokens, whether they are carried as discrete indices or as continuous embeddings.

\subsubsection{Token Vocabulary Standardization} 

Different LMs employ distinct tokenization schemes, for example, byte pair encoding (BPE)~\cite{sennrichNeuralMachineTranslation2016}, WordPiece~\cite{schusterJapaneseKoreanVoice2012}, and SentencePiece~\cite{kudoSentencePieceSimpleLanguage2018}, with varying vocabulary sizes and granularities. This heterogeneity complicates cross-system communication, as tokens from one model may not be directly interpretable by another. Establishing a universal token vocabulary or developing efficient token translation mechanisms is essential for reliable interoperability in heterogeneous TokenCom networks. Tokenizer agreement~\cite{zeinaliWirelessTokenComRLBased2026} offers a third option, standardizing a candidate set of tokenizers and a negotiation protocol rather than the vocabulary itself, so that the pair in use is settled per session and becomes part of resource allocation rather than a fixed constraint.

\subsubsection{Token-Level Quality of Service} 

Traditional quality of service (QoS) mechanisms operate at the bit or packet level, focusing on metrics such as throughput, latency, and packet loss rate. TokenCom requires semantic-aware QoS guarantees that account for the varying importance of different tokens. For instance, tokens representing critical semantic information (e.g., subject or action in a sentence) should receive higher priority than less important tokens (e.g., articles or conjunctions). Designing token-level QoS frameworks that can dynamically allocate resources based on semantic importance remains an open challenge.

\subsubsection{Semantic-Aware Token-Level Metrics} 

The token error rate (TER)\footnote{In~\cite{qiaoToDMALargeModelDriven2025} the token sequence of device $k$ is written as a one-hot matrix $\mathbf{B}_k \in \{0,1\}^{Q \times N}$, with $Q$ the codebook size and $N$ the sequence length, and the TER over $K$ devices is defined as $\frac{1}{2NK} \sum_{k=1}^{K} \lVert \hat{\mathbf{B}}_k - \mathbf{B}_k \rVert_0$. A position decoded as a different index flips two entries of the matrix, so per sequence this equals $\frac{1}{N} \sum_{i=1}^{N} \mathbf{1}\{\hat{t}_i \neq t_i\}$, in which $t_i$ and $\hat{t}_i$ denote the transmitted and the received token at position $i$, and every mismatch enters at unit cost.} used to evaluate TokenCom is the fraction of positions at which the recovered token index differs from the transmitted one~\cite{qiaoToDMALargeModelDriven2025}, so every substitution contributes equally regardless of which token replaced which. The semantic consequences are not equal. In speech recognition, a baseline set of hypotheses and two sets generated from it to hold the word error rate at 7.44\% while changing which words are wrong differ in named entity recognition F1 score from 0.590 to 0.846~\cite{kimSemanticDistanceNew2021}. A comparable spread is expected over tokens, since an error in a background region may be imperceptible while an error in a foreground region may leave the object of interest unrecognizable~\cite{xuDataImportanceAwarePowerAllocation2025}. In related settings, semantic difference has been measured as the distance between sentence embeddings~\cite{kimSemanticDistanceNew2021}, and importance as the sensitivity of a downstream task to a feature channel~\cite{sunTaskOrientedSemanticCommunication2025} or the causal contribution of a token to long-horizon return in closed-loop control~\cite{wangWorldModelEnabledCausal2026}. These factors suggest replacing the unit cost that TER assigns to every mismatch with a weight of the form $w(t_i,\hat{t}_i) = \alpha_i \delta(t_i,\hat{t}_i)$, averaged over the sequence, in which the positional factor $\alpha_i$ measures how strongly the remainder of the sequence depends on the token at position $i$, and the semantic factor $\delta$ grows with the distance between the transmitted and the received token, capturing both their semantic difference and the resulting change in downstream task performance.

\subsubsection{Token Synchronization and Alignment} 

In distributed TokenCom scenarios involving multiple devices or edge nodes, maintaining token-level synchronization is crucial. Misalignment or out-of-order token delivery can severely degrade semantic reconstruction quality. Unlike bit-level synchronization, token synchronization must preserve semantic coherence across potentially variable-length sequences, requiring novel protocols that account for the discrete and context-dependent nature of tokens.

\subsubsection{Token Error Propagation} 

Due to the autoregressive nature of LMs, a single corrupted token can propagate errors throughout the entire generated sequence. This error amplification effect is more severe in TokenCom than in traditional communication, where bit errors are typically localized. The two levels are chained. An undetected bit error can decode to a valid but incorrect index, and the model then conditions the rest of the sequence on that wrong token. Developing robust error detection and correction mechanisms at the token level, possibly leveraging the LM's contextual understanding, is critical for reliable TokenCom deployment. The severity depends in part on the tokenizer's output order: an error in tokens on which many others are conditioned, such as the coarse scales of a multi-scale tokenizer, degrades a correspondingly larger portion of the reconstruction. Conversely, TokenCom admits a form of containment unavailable to bit-level systems, since a receiver holding a generative model can conceal a detected error at the semantic level by regenerating a contextually consistent token. The open problem is then less the correction itself than reliably identifying which tokens are corrupted.

\subsubsection{LM Service-Aware Token Scheduling} 

When TokenCom extends to LM service delivery, which encompasses MoE expert routing, disaggregated prefill-decode serving, and distributed speculative decoding, new scheduling challenges emerge. These include load balancing across expert nodes under dynamic token distributions~\cite{megascaleMoE2025}, minimizing KV cache transfer latency across disaggregated clusters~\cite{mooncake2024,distserve2024}, and coordinating draft-verification token round-trips~\cite{zhengCommunicationEfficientCollaborativeLLM2025}. To date, these mechanisms have been developed either for data-center interconnects or for wireless links whose rate is assumed constant. In TokenCom they must instead operate over time-varying channels, whose achievable rate can change between successive token exchanges. Medium selection inherits this variation. The rate at which cache transfer overtakes token forwarding is fixed by the payload sizes and the receiver compute budget, but the link crosses that rate as the channel varies, which makes the choice a per-exchange decision rather than a per-session configuration. The residual error rate each medium tolerates enters the same decision, and measuring it is a prerequisite for scheduling the two media over a fading link. Unlike semantic QoS, which prioritizes meaning fidelity, LM service-aware scheduling must jointly optimize AI inference correctness, system throughput, and network resource utilization, which constitutes a fundamentally new design space that calls for co-design of communication protocols and LM inference engines.

In summary, TokenCom represents an important step toward unified, intelligent, and semantic-aware communication systems, but practical deployment will require advances in privacy protection, efficiency optimization, token unification, and system robustness, as well as in the token-level challenges specific to TokenCom.

\section{Conclusion}\label{sec:Conclusion} 

This survey reviewed the evolution of LM-driven SemCom toward TokenCom and examined how LMs are reshaping intelligent wireless communication. Existing LM-driven SemCom spans source-centric semantic coding, channel semantics, and collaborative edge-device intelligence, yet its semantic representations remain largely tied to specific modalities, models, and tasks.

TokenCom addresses this fragmentation by adopting tokens, the native processing units of LMs, as a unified interface connecting semantic representation, transmission, and distributed inference. Beyond semantic information delivery, it extends communication design toward LM services and embodied and agentic intelligence. In doing so, it introduces a model-native abstraction above the bit level, where semantic importance, reliability, and resource allocation can be jointly considered, while the bit remains the unit of transport.

Realizing this paradigm still requires advances in privacy and security, efficiency, interoperability, robustness, token-level QoS, error handling, and service-aware scheduling. TokenCom therefore represents not merely the integration of LMs into communication systems, but a step toward the co-design of communication and intelligence for LM-native 6G connectivity.

\bibliographystyle{scis}
\bibliography{refs}

\end{document}